\RequirePackage{fix-cm}
\documentclass[twocolumn]{svjour3}          
\smartqed  
\usepackage{graphicx}
\usepackage{epstopdf}

\usepackage{mathptmx}      
\usepackage{amsmath,amssymb}
\usepackage{siunitx}
\usepackage[dvipsnames]{xcolor}  
\definecolor{linkcolor}{rgb}{0,0,0.42} 
\usepackage[]{natbib}
\usepackage[colorlinks=true,citecolor=linkcolor,urlcolor=linkcolor,linkcolor=linkcolor]{hyperref}
\newcommand\Rey{\mbox{\textit{Re}}}  
\newcommand{\overbar}[1]{\mkern 1.5mu\overline{\mkern-1.5mu#1\mkern-1.5mu}\mkern 1.5mu} 
\makeatletter
\g@addto@macro\@floatboxreset\centering
\makeatother

\begin{document}

\title{Reactive control of the dynamics of a fully turbulent wake using real-time PIV
}


\author{Eliott Varon \and
        Jean-Luc Aider \and
        Yoann Eulalie \and
        Stephie Edwige \and
        Philippe Gilotte.
}


\institute{E. Varon \href{https://orcid.org/0000-0003-4382-2269}{\includegraphics[scale=.5]{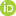}} \and J.-L. Aider \at
              Laboratoire de Physique et M\'ecanique des Milieux h\'et\'erog\`enes (PMMH), CNRS, ESPCI Paris, PSL Research University, 10 rue Vauquelin, Paris, France ; Sorbonne Universit\'e, Univ. Paris Diderot. \\
              \email{jean-luc.aider@espci.fr}           
           \and
           Y. Eulalie \and S. Edwige \and P. Gilotte \at
              Plastic Omnium Intelligent Exterior Systems (POIES), Parc Industriel de la Plaine de l'Ain, Sainte-Julie, France.
}

\date{Received: 03 December Revised : 27 Mai 2019 2018 Accepted: 25 June 2019 \\
This is a pre-print of an article published in Experiments in Fluids. The final authenticated version is available online at: https://doi.org/10.1007/s00348-019-2766-6.}

\maketitle

\begin{abstract}
In this study we focus on the control of the dynamics of 3D turbulent wake downstream a square-back Ahmed body ($\Rey_H=3.9\times10^5$). The peculiar dynamics of such a wake are first characterized through the trajectories of the pressure barycenter over the rear part of the model as well as the recirculation barycenter in the wake. In particular it is shown that these dynamics allow the definition of three different states: the two so-called reflectional symmetry-breaking (RSB) modes and the transient symmetric (TS) mode. It was shown recently that the time-fluctuations of the pressure barycenter could be characterized as a weak chaotic system with a well-defined attractor \citep{Varon2017}. We show that the dynamics of the bimodal wake can then be forced into a stable asymmetric or symmetric state in open loop control, using tangential continuous or pulsed blowing in three different regions along the upper edge of the rear part of the model. Finally, a simple closed-loop opposition control, based on real-time identification of the wake barycenter in the PIV fields, is used to force the chaotic dynamics of the wake into a regular oscillatory motion at a well-controlled frequency. Depending on the actuation parameters, the wake dynamics can also be switched from bimodal to a new multimodal behavior. We show that this new mode also exhibits a peculiar dynamics with an up-down instead of left-right chaotic oscillations. Interestingly, the recirculation area (size of the recirculation bubble) is much more reduced for the closed-loop experiments when the jets are pulsed rather than continuous. For the pulsed jets, the reduction is also increased when the proper frequency is chosen.
\keywords{Wake control \and Optical flow}
\PACS{47.27.Cn} 
\end{abstract}

\section{Introduction}
\label{intro}

The turbulent wakes downstream of three-dimensional (3D) bluff bodies, like cylinders (circular or square), spheres or cubes, either wall-mounted or in the freestream, can be very multifaceted, exhibiting large-scale and small-scale coherent structures, either spanwise or streamwise, with strongly intermittent behaviors  \citep{Williamson1996,Castro1977,Bailey2002,Gumowski2008,Yun2006,Klotz2014,Grandemange2014b}. 

Passenger cars or trucks can be considered as 3D bluff-bodies interacting with a wall (underbody flow). The flow over a real vehicle is very complex but most of the aerodynamics forces are related to various flow separations over the front part (A-pillar vortices, wake of the rear-view mirrors) to the massive flow separation over the rear part, which is responsible of 50 to 70 $\%$ of the drag, depending on the geometry. Studying the structure and dynamics of the 3D wake downstream of the vehicle is then crucial to optimize and reduce the overall drag and related fuel consumption~\citep{Hucho2013}. As a consequence, one can focus on simplified geometries that could mimic the wake of a real vehicle.

In response to the need for a simplified 3D model for automotive aerodynamics studies, the so-called Ahmed body has thus been designed obeying a reflectionnal symmetry \citep{Morel1978,Ahmed1984}. In this geometry, only the rear-slant angle $\alpha$ can be changed ($0^\circ \leqslant \alpha \leqslant 90^\circ$). This simple modification leads to drastic modifications of the overall structure of the wake, together with large variations of the aerodynamic drag coefficient. As an illustration, the slanted configuration ($\alpha=25^\circ$) involves large-scale streamwise longitudinal vortices \citep{Lienhart2002,Beaudoin2004}, together with an unsteady recirculation bubble over the inclined surface \citep{Hinterberger2004,Krajnovic2005}, a pair of a horseshoe vortices containing each a recirculation bubble \citep{Venning2017}, and spanwise Kelvin-Helmholtz vortices, together with smaller-scales turbulent structures. 

The wake behind the square-back configuration ($\alpha=0^\circ$) carries other topologies and dynamics. Indeed, time-averaged velocity and pressure fields reveal a toroidal vortex in the near wake \citep{Duell1999,Krajnovic2003} and spectral analyses highlight a low-frequency mechanism, the bubble pumping \citep{Duell1999,Volpe2015}. Recently, a right-left oscillation of the global wake has been observed, defining the so-called reflectional symmetry breaking (RSB) modes and leading to the bi-stable (or bimodal) wake \citep{Grandemange2013a,Oesth2014}, which appears from the laminar regime \citep{Grandemange2012a,Evstafyeva2017}. This behavior is sensitive to other geometric parameters like the aspect ratio of the bluff-body cross-section and the ground clearance \citep{Grandemange2013c}, and to experimental conditions like the yaw angle \citep{Cadot2015,Volpe2015}. A transient symmetric state is even identified when a cluster-based reduced-order model of the wake is statistically computed \citep{Kaiser2014}.

Our objective in the present study is to implement a closed-loop control based on measurements derived from real-time Particle Image Velocimetry (PIV). Nevertheless, in some cases, controlling the large-scale fluctuations of a turbulent wake is a key step toward the control of various global quantities like the drag coefficient. Actually, stabilization of the wake by suppressing the bimodal behavior is of interest to industrial applications since it can lead to drag reduction \citep{Grandemange2014a,Cadot2015}. Unlike the previous examples on the control of the Lorenz system, this implies here to steady an unstable symmetric state. 

In practical applications the base pressure is often used as the best information to monitor the near wake. But obtaining data directly in the wake allows to better understand its behavior. PIV is well suited for that since it is a non-invasive method. Furthermore, to monitor the bimodal behavior it is possible to relate quantities derived from velocity fields (spanwise component, fluctuations or recirculation intensity barycenter) to quantities derived from base pressure (pressure center or pressure gradient) \citep{Grandemange2013a,Volpe2015,Varon2017}. We need thus to fast compute the velocity fields and their derived quantities.

Real-time PIV has already been reported for FFT-PIV either by reducing the processed images \citep{Siegel2003,Hauet2008,Willert2010} or by using Field-Programmable Gate Arrays (FPGA) \citep{Yu2006,IriarteMunoz2009}, up to 15 fps at $\Rey=1.5 \times 10^4$. Even a closed-loop control using scattered velocity fields from real-time PIV ($\sim 10^2$~vectors at 20~Hz) has previously been successfully applied by \citet{Roberts2012phd} to minimize a pole-cart system plunged into water. Nonetheless, these methods seem unable to provide dense data fields at higher Reynolds numbers. 

As an alternative to the FFT-PIV, three optical flow algorithms have been specifically designed to process images intended for PIV. Based on dynamic programming, the first method computes iteratively the flow displacement over strips of the snapshots \citep{Quenot1998}. It has even been recently applied to stereo-PIV experiments \citep{Faure2010,Douay2013}. The second algorithm derives from the continuity equation and a second-order regularizer \citep{Corpetti2002,Corpetti2006}. Even if these both algorithms give dense velocity fields (one vector per pixel), it seems that their spatial resolution is not better than advanced FFT-PIV \citep{Stanislas2008} and their computation time is at least of the same order as standard FFT-PIV. Following improvements in the displacement estimations of the local approach \citep{LeBesnerais2005}, \citet{Champagnat2011} proposed a faster optical flow algorithm dedicated to PIV application, consisting in estimating at each pixel $m$ the intensity displacement which minimizes the sum of square differences between the intensity over a warped interrogation window (IW) centred in $m$ at time $t$ and the intensity over the warped IW at time $t'=t+dt$. \cite{Davoust2012} and \cite{Sartor2012} used this algorithm to fast post-process their snapshots for a turbulence jet flow study and for an investigation of the interaction between a shock wave and a turbulent boundary layer respectively.

More details about the principle of optical flow computations and a rigorous demonstration of its offline accuracy are given by \citet{Champagnat2011}, \citet{Pan2015} and \citet{Plyer2016}. One of the features of this algorithm is its scalability which makes it very efficient on the Graphics Processor Unit (GPU) architecture. Another feature is its ability to handle regions with large velocity gradients and also to lead to a dense vector field (one vector per pixel). Gautier and Aider \citep{Gautier2014,Gautier2015a,Gautier2015b,Gautier2015c} achieved the implementation of the algorithm on a GPU to enable for the first-time \emph{real-time} high-frequency computations of PIV fields ($\sim 10^6$~vectors at 50~Hz) in closed-loop flow control experiments of a backward-facing step flow in an hydrodynamic channel. 

Recently, it was shown that the large-scale slow-time oscillations behaves like a high dimensional chaotic system with a well-defined strange attractor \citep{Varon2017}. This result was obtained through the analysis of the time-series of the coordinate of the barycenter of the wall-pressure field measured over the rear vertical part of the model. It was the first clear demonstration that the global dynamics of a turbulent 3D wake can be reduced to a simple dynamic system characterized by a set of two variables. The same kind of results was recently found in a turbulent swirling flow \citep{Faranda2017}. If such a simplification is possible, then it should also be possible to apply simple forcing to control the overall dynamics of the 3D wake. This is precisely the objective of the present study.

Previous experimental studies show how different control strategies change the wake dynamics behind a square-back bluff body. For instance, such a wake has been stabilized toward a symmetric state by adding a well-designed cavity behind the rear surface \citep{Evrard2016}. Using pulsed jets along continuous slits at the trailing edges, the unsteady dynamics of the wake can be amplified by selecting either the vortex shedding frequency for an asymmetric forcing configuration, or its subharmonic for a symmetric forcing configuration \citep{Barros2016a}. Inversely, selecting higher actuating frequencies damps the dynamics \citep{Barros2016b}. Feedback controls achieve the suppression of the RSB modes, based either on a linearized stochastic model of the wake dynamics using flaps \citep{Brackston2016}, or on an opposition strategy using lateral slit jets \citep{Li2016}. In the present study, we define a similar control law as the latter one but the sensors being the velocity fields and the actuators being discontinuous jets only localized at the top trailing edge. To our knowledge, this is the first tentative to control a fully turbulent 3D wake through real-time PIV.

The rest of the paper is organized as follows. First, the experimental setup and methods are presented in \autoref{sec:setup}. Then, the natural, uncontrolled flow is briefly recalled in \autoref{sec:natural}. The open-loop forcing of the wake is then presented in \autoref{sec:opencontrol}, showing that the bimodal state can be forced into a single mode state, either symmetric or asymmetric. Finally, \autoref{sec:closedcontrol} presents the closed-loop experiments showing that it is possible to drive the random dynamics into a regular time-periodic system. 

\section{Experimental setup}
\label{sec:setup}

	\begin{figure*}
	\includegraphics{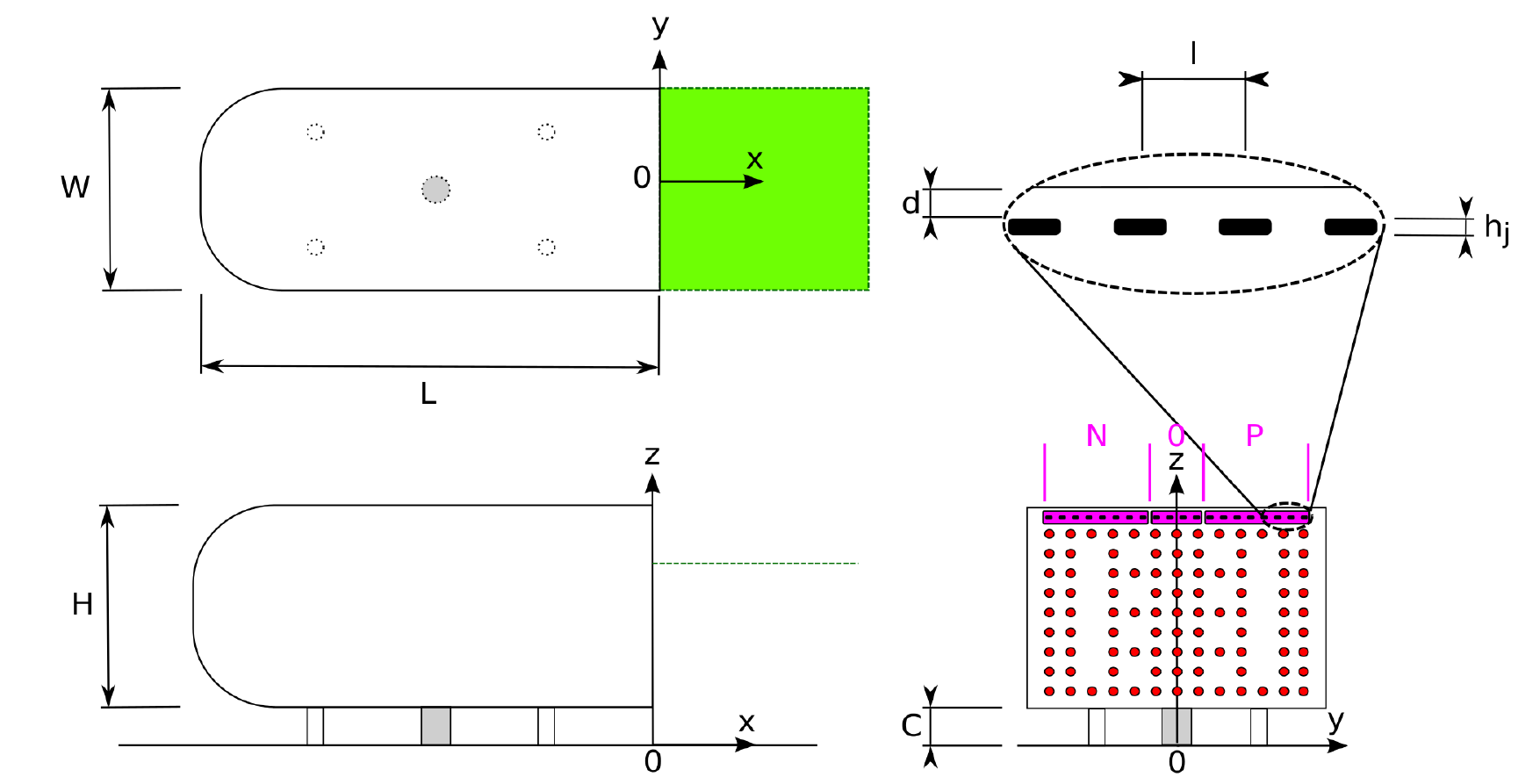}
	\caption{\label{fig:Ahmed_control_sketch}Upper view (top left figure), side view (lower left figure) and view from behind (lower right figure) of the Ahmed body. The main horizontal PIV measurement plane is shown on top (green rectangle) and lower left (green line) figures. The rear part of the model is mapped with 95 pressure sensors (red circles on lower right figure). The jets outputs are gathered in three groups (pink): ``N'', ``0'' and ``P'', corresponding respectively to actuations in the left, central or right upper regions. The jets are supplied through pneumatic tubes passing inside a central pipe (grey part on lower figures).}
	\end{figure*}

\subsection{Ahmed Body}

The bluff body is a 0.7 scale of the original Ahmed body \cite{Ahmed1984}: it is $L=0.731$ m long, $H=0.202$ m high and $W=0.272$ m wide, as described in \citet{Varon2017}. The rear part of the model is a square-back geometry with sharp edges leading to a massive separation of the flow and to the creation of a large 3D recirculation bubble over the rear part of the model.

\subsection{Wind-tunnel}

Experiments have been carried out in the subsonic ``Lucien Malavard" closed wind tunnel of the PRISME laboratory (Orl\'eans, France). The test section of this wind tunnel is 5 m long, with a square cross-section 2~m by 2~m wide, resulting in a  blockage ratio lower than $3\%$. The model is mounted on a raised floor with a properly profiled leading edge and an adjustable trailing edge to avoid undesired flow separations. The ground clearance is set to $C/H=0.248$ to ensure the lateral instability \cite{Grandemange2013c}. In the following, the free-stream velocity is $U_{\infty} =30$~m.s\textsuperscript{-1}, which corresponds to a Reynolds number based on the height of the model $\Rey_H={U_{\infty}H}/{\nu_{air}} = 3.9 \times 10^5$, where $\nu_{air}$ is the air kinematic viscosity at ambient temperature. The free-stream turbulence level is lower than $0.4\%$.
The origin of the $(x, y, z)$ coordinate axis (Fig.~\ref{fig:Ahmed_control_sketch}) is located on the rear of the model ($x=0$), in the vertical symmetry plane ($y=0$) and on the raised floor ($z=0$). Nondimensionalization is applied to spatial variables such as $x^*=x/H$, $y^*=y/H$, $z^*=z/H$, while the convective time and the Strouhal number are respectively defined as $t^* = tU_\infty/H$ and $St_H = fH/U_\infty$ .

\subsection{Sensors}
\label{sensors}

\subsubsection{Wall pressure measurements}

The wall-pressure distribution over the rear part of the model is studied using a set of 95 pressure probes defining a measurement area denoted $S_p$ and covering 70$\%$ of the entire rear surface $S_r$, as shown on Fig.~\ref{fig:Ahmed_control_sketch}. Each vinyl is 2~$cm$ away from each of its neighbors and is connected to a 32-channels \textit{microDAQ} pressure scanner (Chell Instruments), ensuring an accuracy of $\pm17$ Pa and located inside the bluff body. The maximum number of sampling points is $3\times10^4$ for each time-series because of hardware limitations. As a consequence, the sampling frequency for the pressure acquisition $f_{P}$ depends on the acquisition time $T_{P}$: {$f_{P}=3\times10^4/T_{P}$}, but cannot exceed 500~$Hz$.

From these measurements of the pressure $p$, we compute the pressure coefficient:
	\begin{equation}
	C_p(t) = \frac{p(t)-p_\infty}{\frac{1}{2} \rho_{air} U_{\infty}^2},
	\end{equation}
where $p_\infty$ is the free-stream static pressure, measured by a Prandtl tube located above the leading edge of the raised floor, and $\rho_{air}$ is the air density. 

\subsubsection{Real-Time Velocity fields measurements}

	\begin{figure*}
	\includegraphics[scale=1]{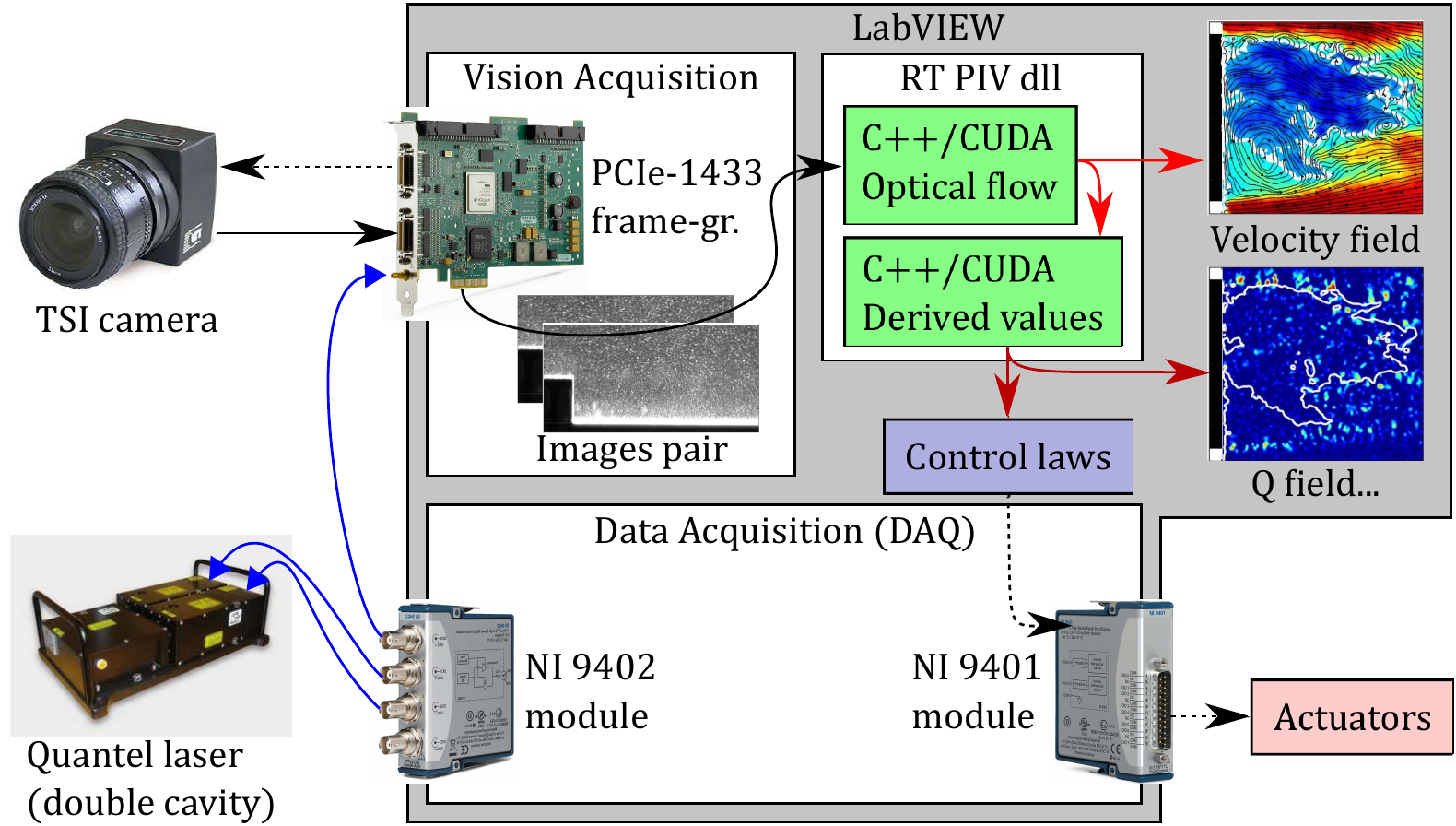}
	\caption{\label{fig:labview}Sketch of the LabVIEW program organization regarding the different devices related to the present wind-tunnel experiment. Solid blue arrows are trig (TTL) signals, dashed black arrows are command signals, solid black arrows are pairs of snapshots, light red arrows are velocity fields and dark red arrow are values derived from the velocity (recirculation area overlying a Q-criterion for instance).}
	\end{figure*}


The two-dimensional two-components (2D-2C) velocity fields are obtained using a standard 7 Hz PIV setup driven by a homemade software. A 200~mJ double-cavity pulsed YaG laser \textit{Twins BSL} (Quantel) is used to generate the laser sheet which enlightens the seeding oil droplets which are injected in the wind-tunnel. A double-frame camera 4 Mp \textit{PowerView Plus 4MP} (TSI) captures the instantaneous pairs of snapshots which are then streamed to the frame-grabber device \textit{NI PCIe-1433} (National Instruments) through Camera Link (CL) cables (510 to 600 Mb.s$^{-1}$). The camera is synchronized to the double cavity laser by programming a \textit{NI-9402} digital module linked to both devices via Bayonet Neill–Concelman (BNC) connectors. 


Based on previous works dedicated to PIV in a hydrodynamics channel \citep{Gautier2014}, the new homemade LabVIEW (National Instruments, USA) interface has been developed to manage the hardware specific to wind-tunnel experiments, together with the optical flow algorithm as sketched in Fig.~\ref{fig:labview}. The entire algorithm is composed of CUDA and C++ functions, all included in a C++ dynamic link library (dll) called RT PIV. The CUDA functions are implemented on the GPU card \textit{GeForce GTX580} \citep{NVIDIAFERMI}. The Vision Acquisition LabVIEW program sends the received images pairs to the dll functions to obtain velocity fields as well as their derived values \emph{in real-time}. These values are then used in a control law to command the actuators driven by the DAQ \textit{NI-9401} digital module.

Regarding the optical flow settings, the IW size is $16 \times 16$ pixels and the calculation is based on three iterations for each of the three pyramid reduction levels. As long as the velocity fields and related quantities are computed and used in a closed-loop control, the acquisition frequency of the 2D-2C PIV velocity fields can reach the maximum camera double-frame rate: $f_{PIV}=7$~Hz. When the data are stored, $f_{PIV}$ drops at $4$~Hz because of the writing time on the solid state drive. With a spatial resolution of 0.21 mm/vector, the investigated PIV plane is the horizontal plane at $z^*=1.00$ as shown in Fig.~\ref{fig:Ahmed_control_sketch}.

\subsubsection{Measured output for the closed-loop control}

As we are interested in controlling the large-scale dynamics of the wake, a global indicator of the state of the wake can be inferred from the instantaneous pressure and velocity fields. We thus use the instantaneous wall pressure barycenter $G_p(x^*=0, y^*, z^*, t)$ and the instantaneous recirculation intensity barycenter $G_{rec}(x^*, y^*, z^*=1, t)$, as defined and illustrated in \citet{Varon2017}. $G_p(t)$ is based on $C_p(t)$, while $G_{rec}$ is related to the recirculation area (negative streamwise velocity) detected in the PIV-plane. Pressure data will be used to illustrate most of our statistic and dynamic results due to their better temporal resolution.

\subsection{The micro-jets}
\label{actuators}

To realize the present control, the rear part of the Ahmed body used in \citet{Varon2017} has been re-designed to include several independent groups of actuators. The bluff-body wake is indeed forced using micro-jets which are distributed along the rear body width, 10~mm under the upper trailing edge ($z^*=1.25$). The jets are either continuous or pulsed at the jet frequency $f_{j}$ using three solenoid valves (Matrix, Italy) integrated inside the model just upstream the blowing jets. Each solenoid valve controls a group of jets, which corresponds to the three regions of actuation defined in Fig.~\ref{fig:Ahmed_control_sketch}.

The pressurized air is supplied through pneumatic tubes passing inside a cylindrical pipe of diameter 32 mm at the center of the bottom face (see the gray part in Fig.~\ref{fig:Ahmed_control_sketch}). Even if using this pipe was unfortunately forced by the bluff body assembly, the lateral instability observed without it \citep{Grandemange2013c} is expected to still exist as observed by \citet{Varon2017} and \citet{Barros2017}.

The outlets of the micro-jets have a rectangular cross-section $l_j \times h_j= 5.5 \times 0.5$~mm\textsuperscript{2}. The distance between each jet is $l = 12.5$~mm. The jets angle $\theta$ relative to the horizontal axis can be varied but in the following it will be fixed to $\theta = 0^o$ (blowing parallel to the freestream velocity). Based on the Strouhal number, the normalized jet frequency is defined as 
	\begin{equation}
	f_{j}^* = \frac{f_{j} H}{U_\infty}.
	\end{equation}
The momentum coefficient is the ratio between the momentum injected by the actuator and the one due to the bluff body:
	\begin{equation}
	c_\mu = 2\frac{l_j h_j \overbar{u_j^2}}{S_r U_\infty^2}.
	\end{equation}
where $\overbar{u_j}$ is the mean jet velocity. For each jet, it is evaluated as $c_\mu=0.053\%$ for continuous blowing and $c_\mu=0.013\%$ for pulsed blowing. It is stressed that the influence of the jet amplitude is not investigated in the present study.

The three regions of actuations are called ``N'', ``0'' and ``P'' (see Fig.~\ref{fig:Ahmed_control_sketch}). ``N'' contains the eight jets located in the Negative part of the $y$-axis, whereas ``P'' contains the eight jets located in the Positive part and ``0'' contains the four central jets (for a total of 20 jets). Due to the temporal constraints imposed for experiments carried in wind tunnel, all recorded control runs lasted only $T^*=8.9 \times 10^3$. Regarding the aerodynamics, only the changes in the pressure coefficient will be considered to analyze influences of the control on the wake to avoid the structural response visible in the aerodynamics measurements.
\section{Base flow}
\label{sec:natural}

The unforced flow is largely investigated in \citet{Varon2017}. The most useful results for the present study are recalled here. the dynamics of the large-scale structures can be tracked either by the pressure center or by the recirculation barycenter since both are highly anti-correlated: when the pressure barycenter is on one side, then the recirculation barycenter is on the other side.

\subsection{Bimodality}

	\begin{figure*}
	\includegraphics[scale=1]{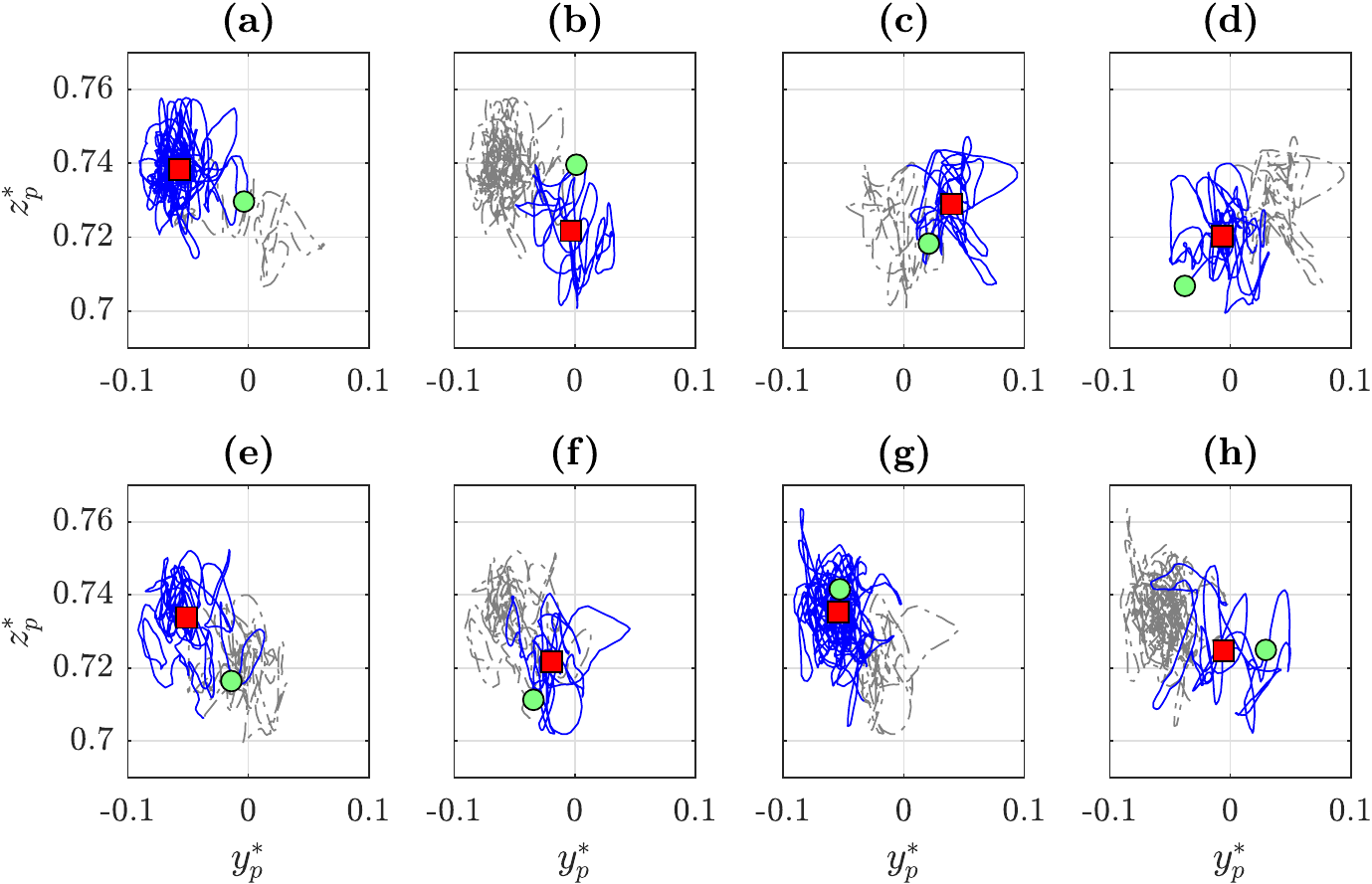}
	\caption{\label{fig:successiveGp_ref}Successive trajectories of $G_{p}(y_p^*, z_p^*)$ during periods of (a) $\Delta t^* = 367$, (b) $\Delta t^* = 105$, (c) $\Delta t^* = 126$, (d) $\Delta t^* = 145$, (e) $\Delta t^* = 194$, (f) $\Delta t^* = 111$, (g) $\Delta t^* = 414$, and (h) $\Delta t^* = 96$. The solid blue line shows the barycenter trajectory and the grey dashed line displays its previous one. The green circle and the red square are respectively the last position and the most probable location in the corresponding sub-figure of $G_{p}$. Animated trajectory of $G_p$ is available in Online Resource 1.}
	\end{figure*}

An important feature of the wake dynamics is its bimodal behavior which is well illustrated in Figs.~\ref{fig:successiveGp_ref} by the evolution of the pressure barycenter $G_p(y_p^*, z_p^*)$ on the rear surface. The barycenter switches from one side to the other ($y_p^*$ switches from negative to positive values) along the spanwise direction while it fluctuates around a mean position along the vertical axis. The most probable location (red square plots) near which $G_p$ evolves during each stay alternates between two opposite spanwise positions, keeping the same vertical position [Figs.~\ref{fig:successiveGp_ref}(a,c and e)]. 

	\begin{figure*}
	\includegraphics[scale=1]{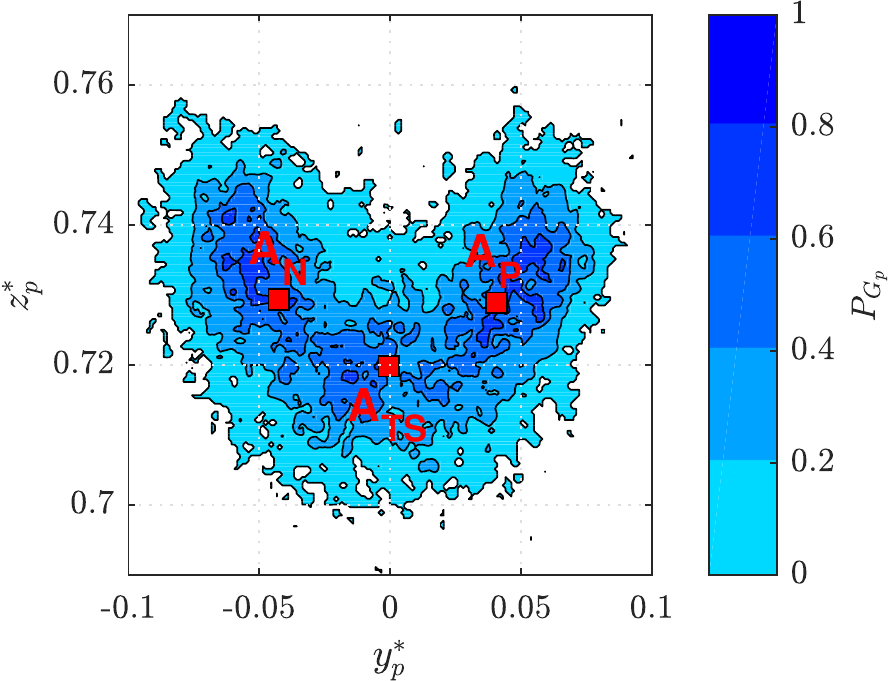}
	\caption{\label{fig:Gp_PDF_ref} Normalized joint PDF of $G_{p}(y_p^*, z_p^*)$ for a 2-min run, together with the conditional-mean positions of $G_{p}$ (red squares) for each identified mode.}
	\end{figure*}

The joint probability density function (PDF) of the pressure barycenter, noted $P_{G_p}$, is displayed in Fig.~\ref{fig:Gp_PDF_ref}. The barycenter moves from one side to the other, between two more probable locations, one located in the negative part of the $y^*$-axis and the other one in the positive part. This is the kind of sidewise bimodal configurations that is indeed expected for a square-back bluff body with such geometric parameters \citep{Grandemange2013c}. Therefore, albeit the central pipe across the under-body flow modifies clearly the near wake topology, particularly in the median longitudinal plan \citep{Lahaye2014,Barros2017}, the bimodal behavior in the spanwise direction is observed like for wakes without the influence of such a pipe \citep{Grandemange2013a,Volpe2015}.

To travel from left [Fig.~\ref{fig:successiveGp_ref}(a)] to right [Fig.~\ref{fig:successiveGp_ref}(c)], the pressure barycenter crosses a central region which can be considered as a third unstable fixed point [Fig.~\ref{fig:successiveGp_ref}(b)]. Even if time spent in one of the sides randomly varies \citep{Grandemange2013a,Brackston2016,Barros2017}, the low-frequency dynamics of the pressure barycenter are weakly chaotic and the two more probable spanwise locations can indeed be considered as the focii of a strange attractor \cite{Varon2017}. Since the complex dynamics of this 3D wake can be reduced to a 2D high-dimensional chaotic system, one should be able to force the system and control its dynamics using the same kind of state parameters.

\subsection{Spectral analysis}

	\begin{figure*}
	\includegraphics{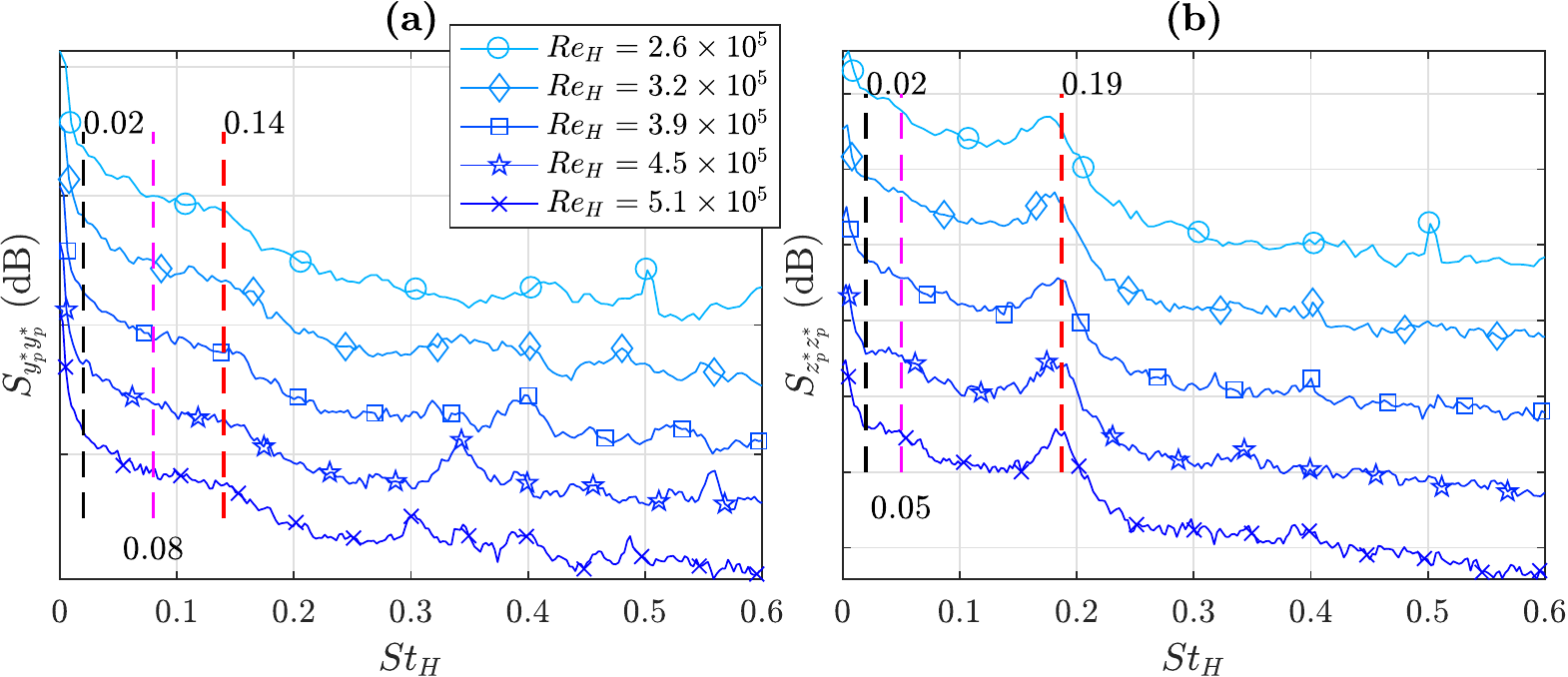}
	\caption{\label{fig:Sxx_yzP_ReH}PSD of the fluctuations of (a) $y_p^*$ and (b) $z_p^*$ for five Reynolds number: $2.6\times 10^5$ (circle), $3.2\times 10^5$ (diamond), $3.9\times 10^5$ (square), $4.5\times 10^5$ (star) and $5.1\times 10^5$ (cross). Curves are shifted for clarity. The frequency resolution is $\Delta St_H = 2 \times 10^{-3}$.}
	\end{figure*}

For different Reynolds numbers, the power spectra density (PSD) of the sidewise position $y_p^*$ [Fig.~\ref{fig:Sxx_yzP_ReH}(a)] and the vertical position $z_p^*$, noted $S_{y_p^*y_p^*}$ and $S_{z_p^*z_p^*}$ respectively, contain high amplitude at a very-low-frequency range ($St_H < 0.02$). The observation of high energy level in this part of the spectrum is common in measurements close to the rear body \citep{Grandemange2013a,Volpe2015}. Considering the average time spent in a RSB mode ($\sim 7 \times 10^2$), it may be associated with the lateral bimodal behavior. 
A decaying but still high level of energy is then observed in $S_{y_p^*y_p^*}$ for $St_H \in \lbrack 0.02, 0.08 \rbrack$. \citet{Khalighi2012}, \citet{Volpe2015} and \citet{McArthur2016} linked  $St_H = 0.08$ to the bubble pumping phenomenon. Then a plateau appears until $St_H \sim 0.14$. It contains the value 0.13 previously found with hot-wire measurements in the lateral shear layers, indicator of vortex shedding \citep{Lahaye2014,Eulalie2014phd,Volpe2015}.
Recent frequency measurements in a simulated turbulent wake behind a circular disk confirm the presence of two close low frequencies: 0.02 and 0.03 \citep{Yang2015}. The lowest one is associated with the asymmetric changes in the wake orientation at such low frequency, whereas the highest one is related to the bubble pumping. Similar results have been obtained for a turbulent axisymmetric body wake \citep{Gentile2016}. Through a POD analysis of the velocity fluctuations, they find very-low-frequency peaks, $St_H \sim 10^{-4}$ and $St_H=10^{-3}$, in the dynamics of the modes associated respectively to the symmetry breaking process and to the bubble pumping. 
The PSD of the vertical position $z_p^*$, $S_{z_p^*z_p^*}$, alone reveals a peak at $St_H = 0.19$, independent from the Reynolds number, as displayed in Fig.~\ref{fig:Sxx_yzP_ReH}(b). This Strouhal number is characteristic of the signature of vortex shedding from the top-bottom shear layers interaction \citep{Parezanovic2012,Lahaye2014,Volpe2015,Barros2016a}, meaning that $z_p^*$ is influenced by these far wake dynamics. This natural feedback inside the wake has been reported by \citet{Broze1994} for a turbulent axisymmetric jet, where the spectral signature of vortex pairings downstream of a nozzle appears in the velocity fluctuations measured at the nozzle exit. For the turbulent axisymmetric body wake, the characteristic vortex shedding frequency also appears in the spectrum of the POD mode describing the switching behavior \citep{Gentile2016}. Physically, the feedback is due to upstream propagating pressure fluctuations.

Thus, the dynamics of the depression seem to be rather dominated by the spanwise bimodal behavior of the recirculation and the top-bottom shear layers interaction. The central pipe may partly intensify this interaction, as highlighted by \citet{Lahaye2014}, but most of the added structures in the underbody flow have higher frequency signatures like the Karman vortex street at $St_H=1$ and $St_H=1.6$ \citep{Eulalie2014phd}. It shall be recalled that the switching process is linked to the turbulent perturbations growth \citep{Brackston2016}.

\subsection{Transient symmetric state}

During most of the flipping processes, the pressure center goes around a central position  $y_p^* \sim 0$, as shown in Figs.~\ref{fig:successiveGp_ref}(b, d, f and h). In average, this step lasts for a much shorter time than the RSB modes. It shall be noteworthy that, after being in this transient state, the wake may also go back in the previous RSB mode [Figs.~\ref{fig:successiveGp_ref}(e-g)].

As mentioned previously, three states can be considered for the system: two corresponding to the RSB modes, and a third one being a transient mode. For a more quantitative analysis, two thresholds for $y_p^*(t)$ has consequently been defined. Above a given positive threshold $y_{p,P}^*$, the wake is in the P mode, and lower than a given negative threshold $y_{p,N}^*$, it is in the N mode. Between these values, it is the transient state. The thresholds are computed as
	\begin{equation}
	y_{p,N}^* = w_{th} \times {y_p^*}_{\max(P_{y_p^*\vert y_p^*<\overbar{y_p^*} })}
	~\textrm{and}~
	y_{p,P}^* = w_{th} \times {y_p^*}_{\max(P_{y_p^* \vert y_p^*>\overbar{y_p^*} })},
	\label{eq:thresholds}
	\end{equation}
where $P_X$ is the PDF of the variable $X$, $\overbar{X}$ is the temporal mean of $X$, and $w_{th}$ is a positive weight to avoid too large thresholds ($w_{th}=0.25$ is chosen). From now on, the state of the wake is defined as negative (N), positive (P) or transient symmetric (TS), where respectively $y_p^*<y_{p,N}^*$, $y_p^*>y_{p,P}^*$ and $y_{p,N}^*<y_p^*<y_{p,P}^*$. The TS state existence has also been recently reported in the wake of similar squareback bluff bodies: through a cluster-based approach \citep{Kaiser2014}, a POD analysis \citep{Pavia2017} and partially-averaged Navier–Stokes simulations \citep{Rao2018}.

Even if the switching process seems to randomly occur, some characteristic time scales can be statistically estimated. Hence, the time the wake spent on each mode is evaluated from all runs, giving $T_{RSB}^* \sim 7 \times 10^2$ for each asymmetric mode, whereas the switch itself lasts for $T_{switch}^* \sim 4 \times 10^1$. As expected, these results have a high dispersion caused by the turbulent noise, their respective standard deviations being $\sigma_{T_{RSB}^*} = 10^2$ and $\sigma_{T_{switch}^*} = 10^1$. A mean switching frequency is also computed by counting the switching occurrences during runs: $f_{switch}^* \sim 3.4 \times 10^{-3}$. In probability terms, the RSB modes exist $83.4\%~(\pm 5.3)$ of the time ($41.1\%$ for the $N$ state, and $42.3\%$ for the $P$ one), whereas the wake is in the unstable $TS$ mode during the remaining $16.6\% ~(\pm 1.2)$. This is the reason why the present wake can also be considered thereafter exhibiting a trimodal dynamics.

The joint PDF of the pressure barycenter is computed and shown in Fig.~\ref{fig:Gp_PDF_ref}. It gives a global overview of its most frequent positions on the rear. As revealed by the time series, three main areas corresponding to the wakes states are highlighted, confirming the trimodal behavior. We calculate then the positions of the identified centers of these areas, denoted $A_N (y_{A_N}^*,z_{A_N}^*)$, $A_P (y_{A_P}^*,z_{A_P}^*)$ and $A_{TS} (y_{A_{TS}}^*,z_{A_{TS}}^*)$ for the N, P and TS states respectively, such as
	\begin{equation}
	\overrightarrow{OA_N} = \left(
	\begin{array}{ll}
	 	{y_p^*}_{\vert y_p^*<y_{p,N}^*}\\
	 	{z_p^*}_{\vert y_p^*<y_{p,N}^*}
	\end{array}
	\right)\textrm{,}~
	\overrightarrow{OA_P} = \left(
	\begin{array}{ll}
	 	{y_p^*}_{\vert y_p^*>y_{p,P}^*}\\
	 	{z_p^*}_{\vert y_p^*>y_{p,P}^*}
	\end{array}
	\right) 
	\textrm{and}~
	\overrightarrow{OA_{TS}} = \left(
	\begin{array}{ll}
	 	{y_p^*}_{\vert y_{p,N}^*<y_p^*<y_{p,P}^*}\\
	 	{z_p^*}_{\vert y_{p,N}^*<y_p^*<y_{p,P}^*}
	\end{array}
	\right).
	\label{eq:attractorCenter}
	\end{equation}
For each area a quasi-attractive center can be identified, such as $y_{A_N}^* \sim - y_{A_P}^*$ and $z_{A_N}^* \sim z_{A_P}^*$. During the switch between these two positions, the pressure barycenter follows preferentially a trajectory along a well-defined path.

	\begin{figure*}
	\includegraphics[scale=1]{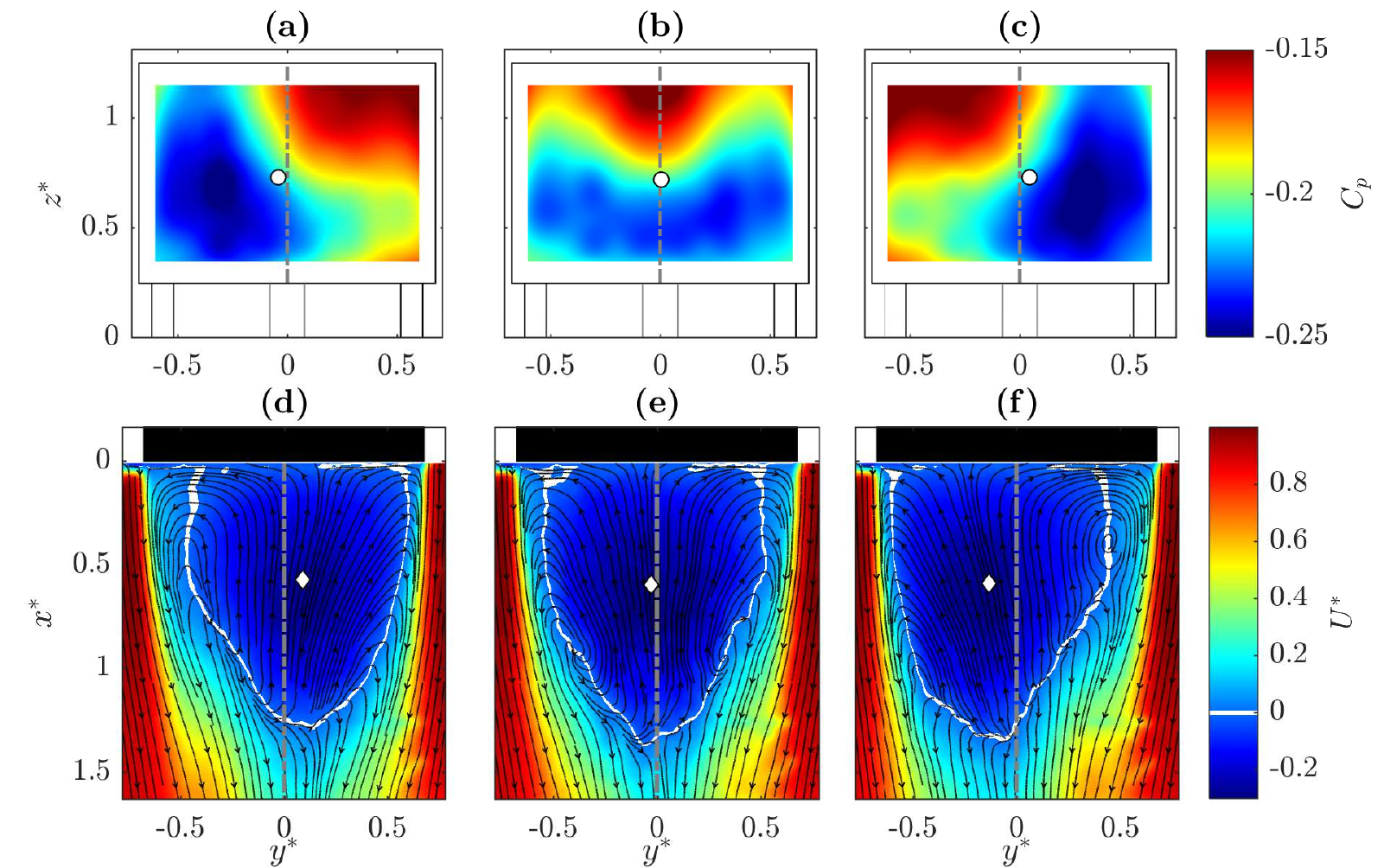}
	\caption{\label{fig:statesNTP_ref}Conditional averages of the pressure coefficient at the rear body and the velocity field at $z^*=1$ for the modes (a-d) N, (b-e) TS and (c-f) P. The average position of the pressure center $G_p$ (white circle) and of the recirculation barycenter $G_{rec}$ (white diamond) are displayed. $y^*=0$ is also displayed (dash-dotted grey lines). Some streamlines are plotted over the streamwise component of the velocity to localize the coherent large-scale structures.}
	\end{figure*}

It is then interesting to visualize the corresponding time-averaged pressure and velocity fields using conditional averaging. Figs.~\ref{fig:statesNTP_ref}(a-c) display the conditional time-averaged pressure fields of the three modes, while Figs.~\ref{fig:statesNTP_ref}(d-f) correspond to their conditional time-averaged velocity fields. The RSB modes presented here have identical pressure and velocity topologies as the ones without considering the transient state \citep{Volpe2015}: the depression is localized on the same side as the vortex being the closest to the rear surface. In the TS state, the wake appears to be a perfect mean of the two RSB modes. Once again, the phase opposition between the pressure and velocity barycenters clearly appears. 

The effects of the modified ground clearance on the wake reversal behind a square-back bluff body have been studied by \citet{Barros2017}. It was reported in particular that the presence of a circular cylinder, very similar to the central pipe in our experiments but closer to the rear part, changes the vertical position of the modes but does not cancel the lateral bimodality.


\section{Open-loop forcing of the wake}
\label{sec:opencontrol}

Three different open-loop (OL) control strategies are investigated to evaluate the forcing on the state of the wake for 1-min runs.  The three open-loop forcings are defined as``OL-P" when only the ``P'' solenoids group is activated, ``OL 0" when only the ``0'' solenoids group is activated and ``OL-P0" when both the ``P'' and ``0'' solenoids groups are activated (Fig.~\ref{fig:Ahmed_control_sketch}). Due to the reflection symmetry the results obtained with the ``P'' group are expected to be the same as for the ``N'' configuration. First we used continuous jets, which means only the forcing region is changed. Then pulsed jets are investigated, especially at the normalized frequency $f_{j}^*=$~0.2. 

\subsection{Continuous blowing in different regions of the upper edge of the rear of the bluff-body}

\subsubsection{Influence on the trajectory of the pressure barycenter}

	\begin{figure*}
	\includegraphics{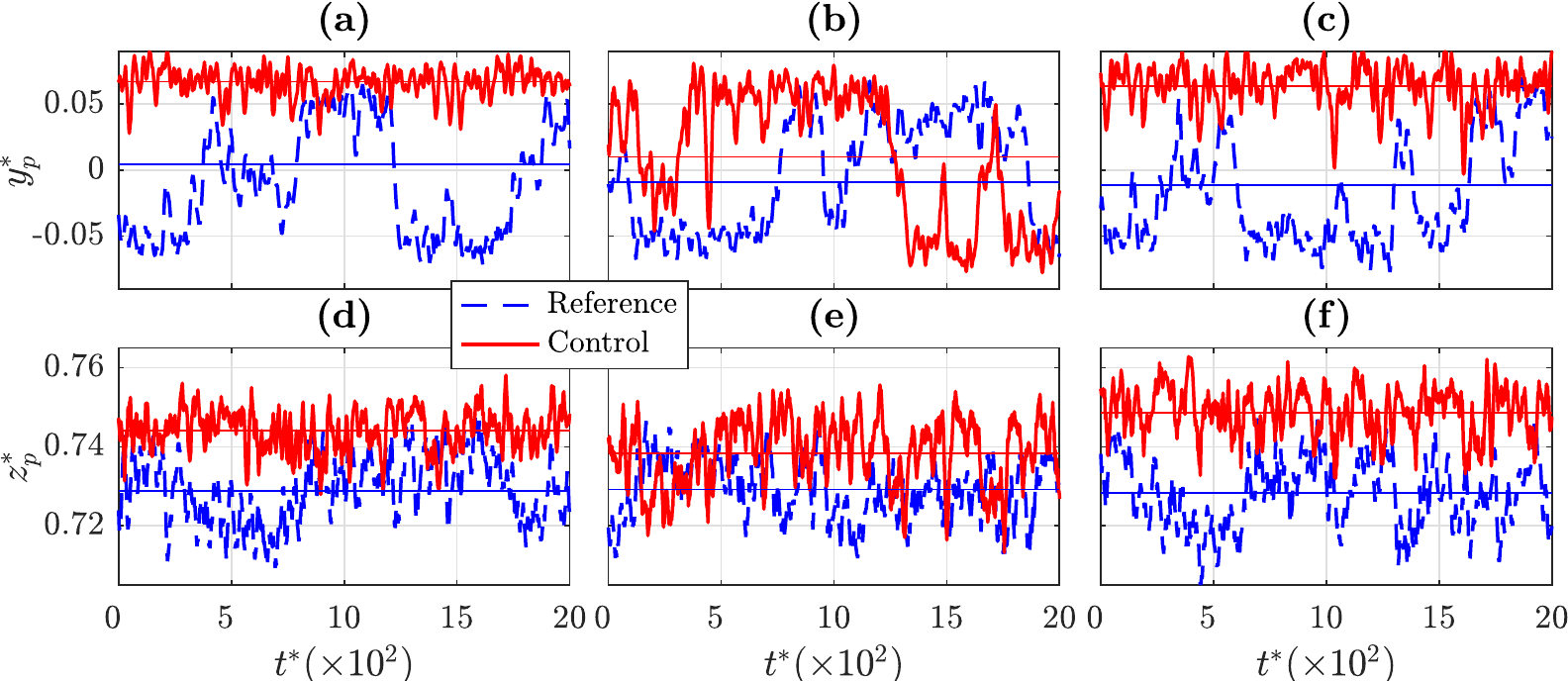}
	\caption{\label{fig:yPzP_ref_BOF0}Excerpts of $y_p^*$ and $y_p^*$ time series for the continuous forcing (a-d) OL-P, (b-e) OL-0, and (c-f) OL-P0 (solid red), all compared to a reference time series (dashed blue). Data are smoothed over $t^*=30$ for clarity.}
	\end{figure*}

The influence of the location of the continuous blowing on the wake dynamics is studied by analysing directly the times series of the pressure center obtained for the three open-loop cases. On one hand, as displayed in Fig.~\ref{fig:yPzP_ref_BOF0}(a) and Fig.~\ref{fig:yPzP_ref_BOF0}(c), blowing from one side drives the depression to remain localized in this side. On the other hand, Fig.~\ref{fig:yPzP_ref_BOF0}(b) reveals that the symmetric forcing seems not to affect the spanwise bimodal behavior. For all the forcing cases, the vertical position of the depression becomes closer to the upper trailing edge. The averaged $z_p^*$ indeed rises $2.1\%$ for OL-P [Fig.~\ref{fig:yPzP_ref_BOF0}(d)], $1.2\%$ for OL-0 [Fig.~\ref{fig:yPzP_ref_BOF0}(e)] and $2.8\%$ for OL-P0 [Fig.~\ref{fig:yPzP_ref_BOF0}(f)].

	\begin{figure*}
	\includegraphics{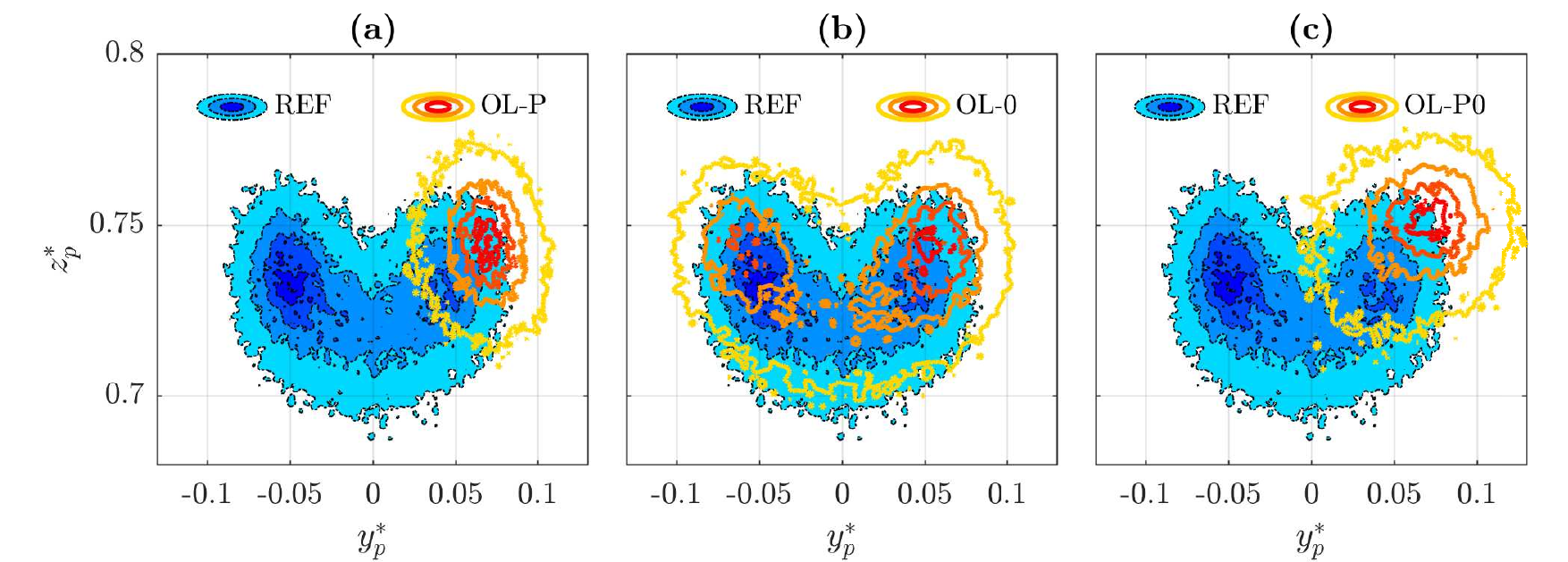}
	\caption{\label{fig:ZpYp_PDF_ref_BOF0}Joint PDF contour of $G_p(y_p^*, z_p^*)$ for the open-loop controls with continuous jets compared to the reference (dashed blue contour): (a) OL-P, (b) OL-0, and (c) OL-P0 (solid red contour). The colormap is the same as Fig.~\ref{fig:Gp_PDF_ref}}
	\end{figure*}

The previous observations are also confirmed by the joint PDFs. Figure~\ref{fig:ZpYp_PDF_ref_BOF0}(a) shows that forcing with continuous jets the upper shear layer on one side of the model (OL-P) strongly stabilizes the wake oscillations. The pressure barycenter is now kept on the side of the blowing ($y_p^* > 0$) and oscillates around a single attractor. If the forcing is extended to the central region (OL-P0), the pressure barycenter is still confined on one side of the model but explores a larger region around the attractor [see Fig.~\ref{fig:ZpYp_PDF_ref_BOF0}(c)]. Finally, forcing the wake only in the central region (OL-0) leads to the same kind of PDF as the natural case [see Fig.~\ref{fig:ZpYp_PDF_ref_BOF0}(b)]. Nevertheless, the probability to remain in the TS state is reduced by half compared to the natural state. The TS mode is thus clearly minimized by the central symmetric continuous blowing.

\subsubsection{Chaotic behavior}

	\begin{figure*}
	\includegraphics{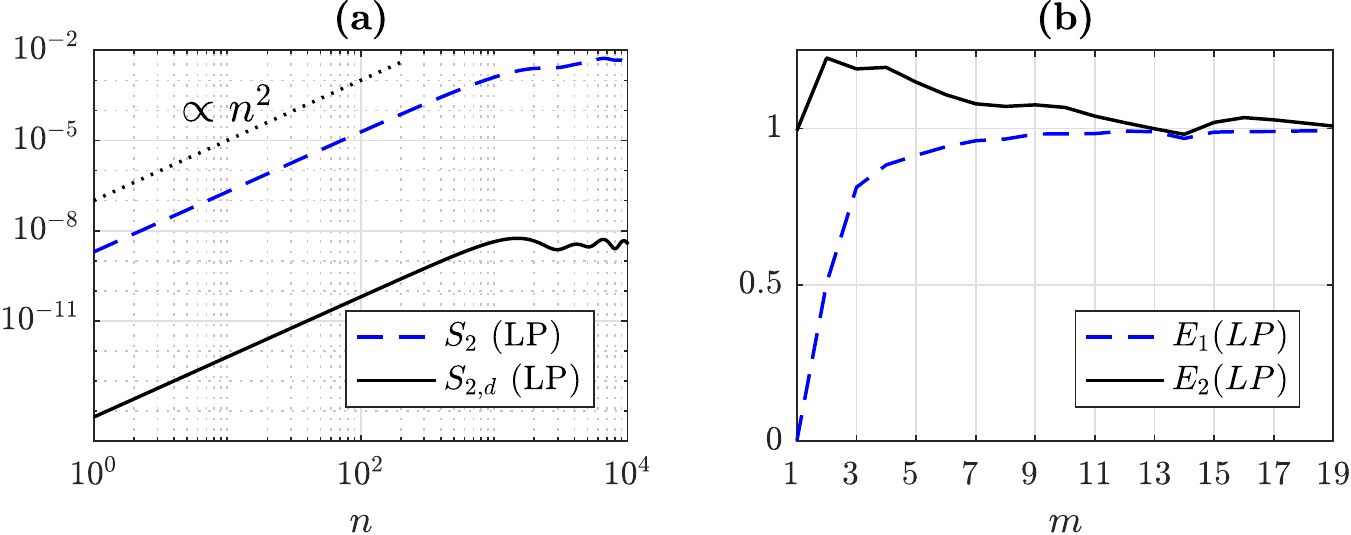}
	\caption{\label{fig:structFunction_embeddingDim_OL0LF}(a) Second-order structure functions of $y_p^*$ (dashed blue line) and of its first derivative (solid black line). (b) $E_1(m)$ (dashed blue line) and $E_2(m)$ (solid black line) computed from $y_p^*$. $f_p=500$~Hz. As explained by\citet{Varon2017}, $y_p^*$ is low-pass (LP) filtered at $f_{LP}^*=7 \times 10^{-3}$ before these calculations.}
	\end{figure*}

Concerning the dynamics of the last case, random switches occurs less often but the weak chaotic behavior in the low-frequency range remains ($St_H < 2 \times 10^{-3}$). First, Fig.~\ref{fig:structFunction_embeddingDim_OL0LF}(a) displays the second order-structure functions of $y_p^*$ and its derivative, denoted $S_2$ and $S_{2,d}$ respectively, measuring the self-affinity of the signal \citep{Provenzale1992}. $S_2$ is defined as
	\begin{equation}
	S_2(n) = \langle \vert y_p^*(i+n)-y_p^*(i) \vert^2 \rangle_i, 
	\end{equation}
where $n$ is the lag and $\langle . \rangle_i$ stands for the average over $N-n$ points. For small $n$, both curves follow a scaling law $n^2$, specific to chaotic dynamics \citep{Mandelbrot1982}. Secondly, Fig.~\ref{fig:structFunction_embeddingDim_OL0LF}(b) shows the computation of the embedding dimension $m$ \citep{Cao1997}. In brief, after building specific state vectors from $y_p^*$ time series for phase space reconstruction, based on the Takens's time-delay embedding method \citep{Takens1981}, the evolution of the mean distance between them for different tested embedding dimensions $m$ is evaluated through a function $E_1$. Then, a minimum embedding dimension $m$ exists if $\forall k>m, E_1(k)=E_1(k+1)$, leading to $m=15$, of the same order as the reference flow \citep{Varon2017}. The same computation is implemented directly with $y_p^*$ and the function is denoted $E_2$. It enables to verify if $y_p^*$ is a deterministic (chaotic) signal if $\exists k \backslash E_2(k) \neq 1$, which is well the case when OL-0 is activated. More informations about the computations are given by \citet{Varon2017}. Unfortunately, according to \citet{Eckmann1986}, the recorded time for these data (1 min) is too short regarding the number of switching events to properly characterize their chaotic dynamics through the calculations of the correlation dimension and the largest Lyapunov exponent.

\subsubsection{Trajectories in phase space}

	\begin{figure*}
	\includegraphics{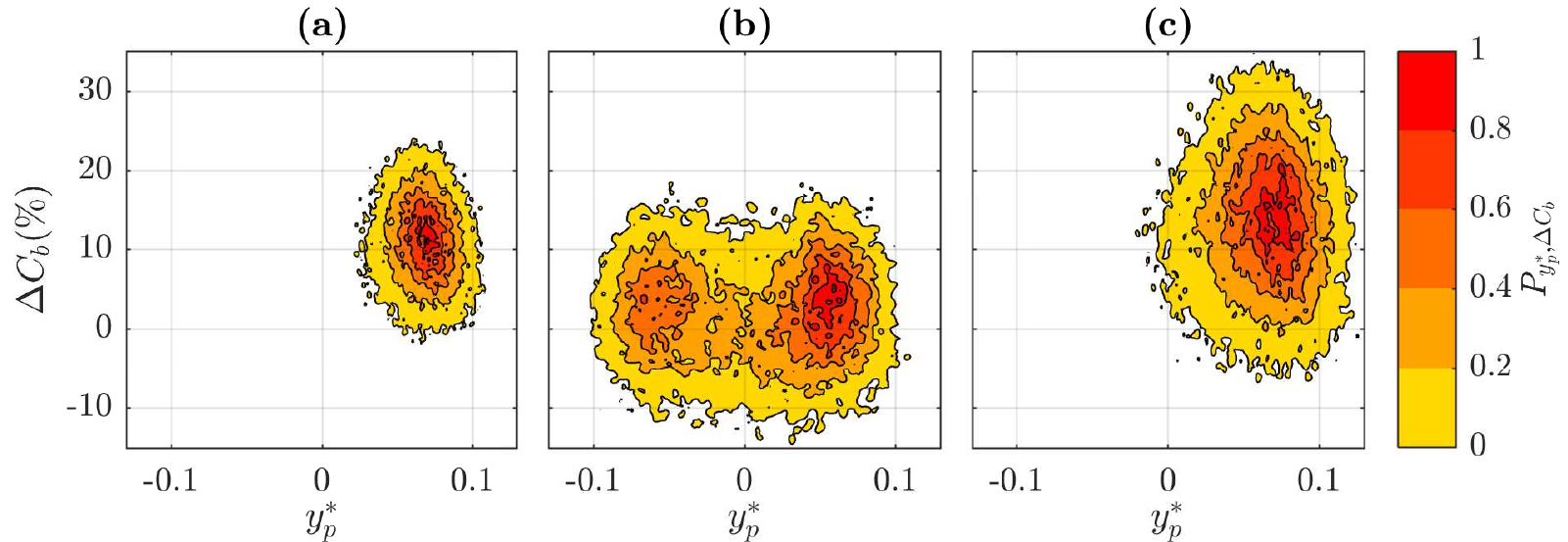}
	\caption{\label{fig:DeltaCbYp_PDF_BO}Joint PDF contour of ($y_p^*$, $\Delta C_b$) for the open-loop controls (a) OL-P, (b) OL-0, and (c) OL-P0 with continuous blowing.}
	\end{figure*}

Instead of looking at the trajectories of the pressure barycenter over the rear of the model in the physical space, it is interesting to look at the trajectory in the phase space. Indeed the state of the wake can be characterized not only by the right-left oscillations (bimodality), through the state parameter $y_p^*$, but also by the normalized fluctuations of the instantaneous base suction coefficient $\Delta{C_b}(t)$
	\begin{equation}
	\Delta{C_b}(t) = \frac{C_b(t) - C_{b,ref}}{C_{b,ref}},
	\label{eq:DeltaCp}
	\end{equation}
where $C_b(t)=-\iint_{S_r}Cp(t)$ and $C_{b,ref}$ is the time-averaged value of $C_b$ for the reference flow. In Fig.~\ref{fig:DeltaCbYp_PDF_BO}(a) and (c), forcing the wake into one of the RSB modes increases roughly by $15 \%$ the depression. These results are well coherent with the sensibility analysis done by \citet{Grandemange2014a}. On the contrary, acting on the symmetric mode of the wake leads to similar value for $C_{b}$ than the reference case [see Fig.~\ref{fig:DeltaCbYp_PDF_BO}(b)]. 

\subsection{Pulsed blowing in different regions of the upper edge of the rear of the bluff-body}

	\begin{figure*}
	\includegraphics{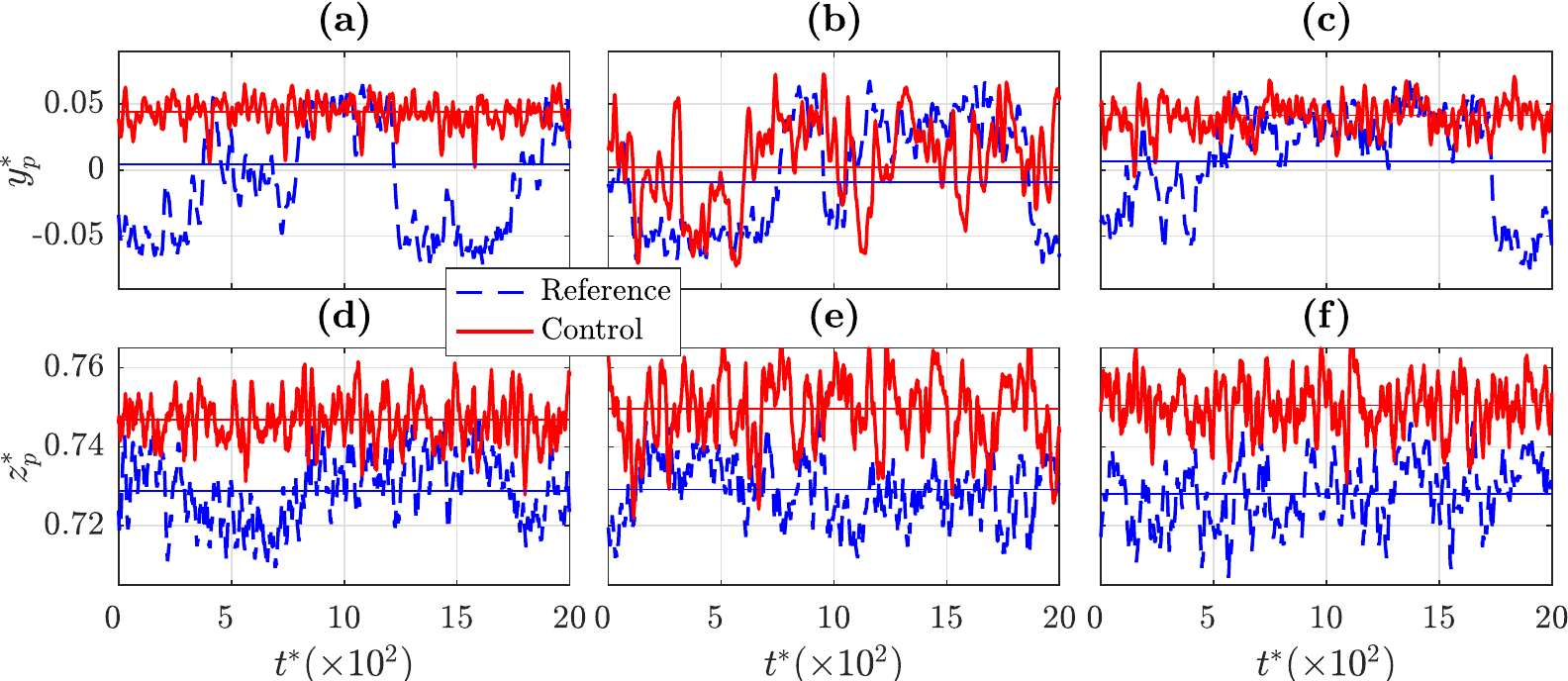}
	\caption{\label{fig:yPzP_ref_BOF30}Excerpts of $z_p^*$ and $z_p^*$ time series for the pulsed jets at $f_{j}^*=0.2$ in configurations (a-d) OL-P, (b-e) OL-0, and (c-f) OL-P0 (solid red), all compared to a reference time series (dashed blue). Data are smoothed over $t^*=30$ for clarity.}
	\end{figure*}

Finally, the previous forcing configuration are investigated using pulsed jets, testing different normalized frequencies $f_{j}$. We present here the results obtained for $f_{j}^*=0.2$ since this almost corresponds to the vortex shedding frequency found in the up-down oscillations of the depression position for the reference case ($f_{j}^*=0.19$ was not properly attainable). At this actuating frequency, the depression behaves almost as previously for the OL-P and OL-P0 cases, but with more regular oscillations in the $z$-direction and a reduction of the spanwise fluctuations, amplifying the decrease in the base pressure. On the contrary, the OL-0 case is strongly modified: the average spanwise position of the depression is now centred [Fig.~\ref{fig:yPzP_ref_BOF30}(b)], the amplitude of the vertical fluctuations doubled while the mean vertical position increases much more than for the continuous blowing [Fig.~\ref{fig:yPzP_ref_BOF30}(e)]. For these reasons, the OL-0 forcing at $f_{j}^*=0.2$ will be more detailed in the following. 

	\begin{figure*}
	\includegraphics{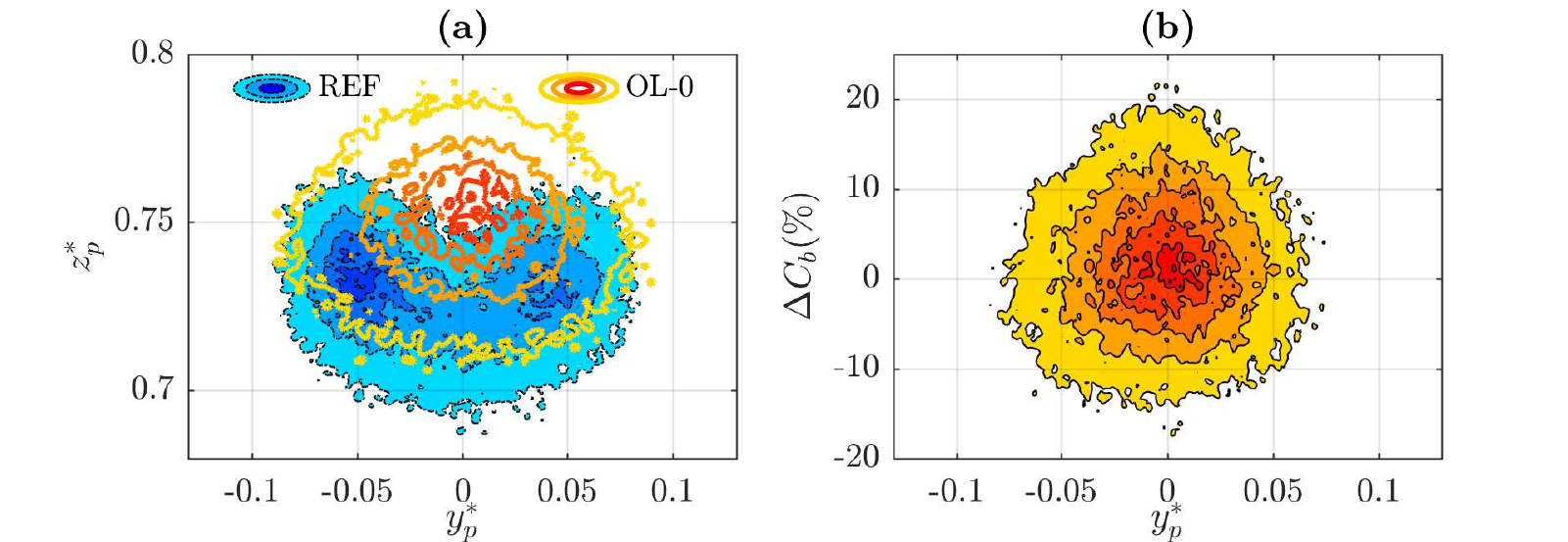}
	\caption{\label{fig:ZpYp_PDF_DeltaCb_BO0F30}Joint PDF contour of (a) $G_p(y_p^*, z_p^*)$ and (b) ($y_p^*$, $\Delta C_b$) with pulsed jets at $f_{j}^*=0.2$. The first joint PDF is compared to the reference (dashed blue contour). Colormaps are the same as Fig.~\ref{fig:ZpYp_PDF_ref_BOF0} and Fig.~\ref{fig:DeltaCbYp_PDF_BO} respectively.}
	\end{figure*}

Actually, the control with pulsed jets at $f_{j}^*=0.2$ leads to a symmetric PDF with a single central attractor for the trajectory of the pressure barycenter, as illustrated by Fig.~\ref{fig:ZpYp_PDF_DeltaCb_BO0F30}(a), exhibiting only one central peak. The probability of being in the TS mode is multiplied by two. This has to be compared to the two attractors trajectories obtained with a constant blowing in the same region [Fig.~\ref{fig:ZpYp_PDF_ref_BOF0}(b)].  

	\begin{figure*}
	\includegraphics[scale=1]{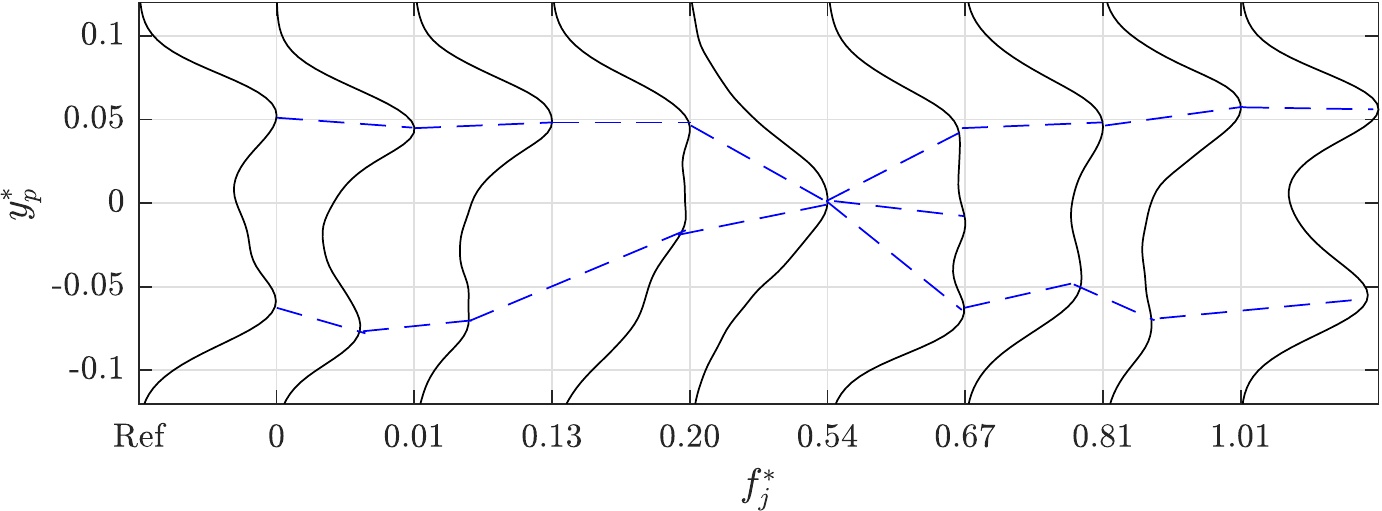}
	\caption{\label{fig:PDF_BO}Normalized PDF for the reference flow (Ref), the continuous blow ($f_j^*=0$) and different actuating frequencies $f_j^*>0$ for OL-0. Curves are shifted according to the respective cases. Forcing the wake with the $f_j^*=0.2$ pulsing frequency clearly leads to a very strong modification of the dynamics of the wake.}
	\end{figure*}

Figure \ref{fig:PDF_BO} stresses the peculiarity of the actuation frequency at $f_{j}^*=0.2$, since no other investigated frequency enables such a symmetrization of the wake. This dynamic response of the wake to a specific perturbation reminds the one obtained for a forced axisymmetric jet resulting to Unstable Periodic Orbits (UPOs) \citep{Broze1994}. As the pressure center of the turbulent wake acts like a strange attractor in the low-frequency range, its space phase shall embed UPOs according to \citet{Auerbach1987}. Therefore, the OL-0 forcing at $f_{j}^*=0.2$ may have enhanced large UPOs \citep{Grebogi1988}, such as the TS mode dominates. Further investigations to detect and map the UPOs from $y_p^*$ \citep{Ma2013} are needed to verify this dynamical interpretation which remains an open question.

The sensitivity analysis made by \citet{Grandemange2014a} shows that targeting the center of the top or of the bottom shear layer with a control cylinder stabilizes the wake in a symmetric mode. In our case, this wake topology is obtained by forcing the top shear layer at the vortex shedding frequency. As highlighted recently, using this frequency for the actuation frequency leads to a strong increase of the pressure drag (over $+25 \%$) by enhancing the entrainment rate of the wake \cite{Barros2016a,Barros2016b}. Thereby, forcing the wake in its center at one of its main resonant frequency ($St_H = 0.2$) induces two opposite effects on the base pressure, resulting finally to its very slight increase ($+0.75 \%$ in average), as shown by Fig.~\ref{fig:ZpYp_PDF_DeltaCb_BO0F30}(b). The impact of a low pulse frequency previously observed may be attenuated in our case as the exit slit of the jets is discontinuous and only localized in the middle of the upper rear part, unlike the forcing in Refs.~\citep{Barros2016a,Barros2016b}. It is noteworthy that base pressure recovery was also obtained at similar and lower normalized jet frequencies for other bluff body wakes \citep{Brunn2006,Joseph2012}, axisymmetric wakes \citep{Sigurdson1995}, and 2D wakes \citep{Pastoor2008}.

\section{Closed-loop control of the wake dynamics}
\label{sec:closedcontrol}

\subsection{Control law}

	\begin{figure*}
	\includegraphics[scale=1]{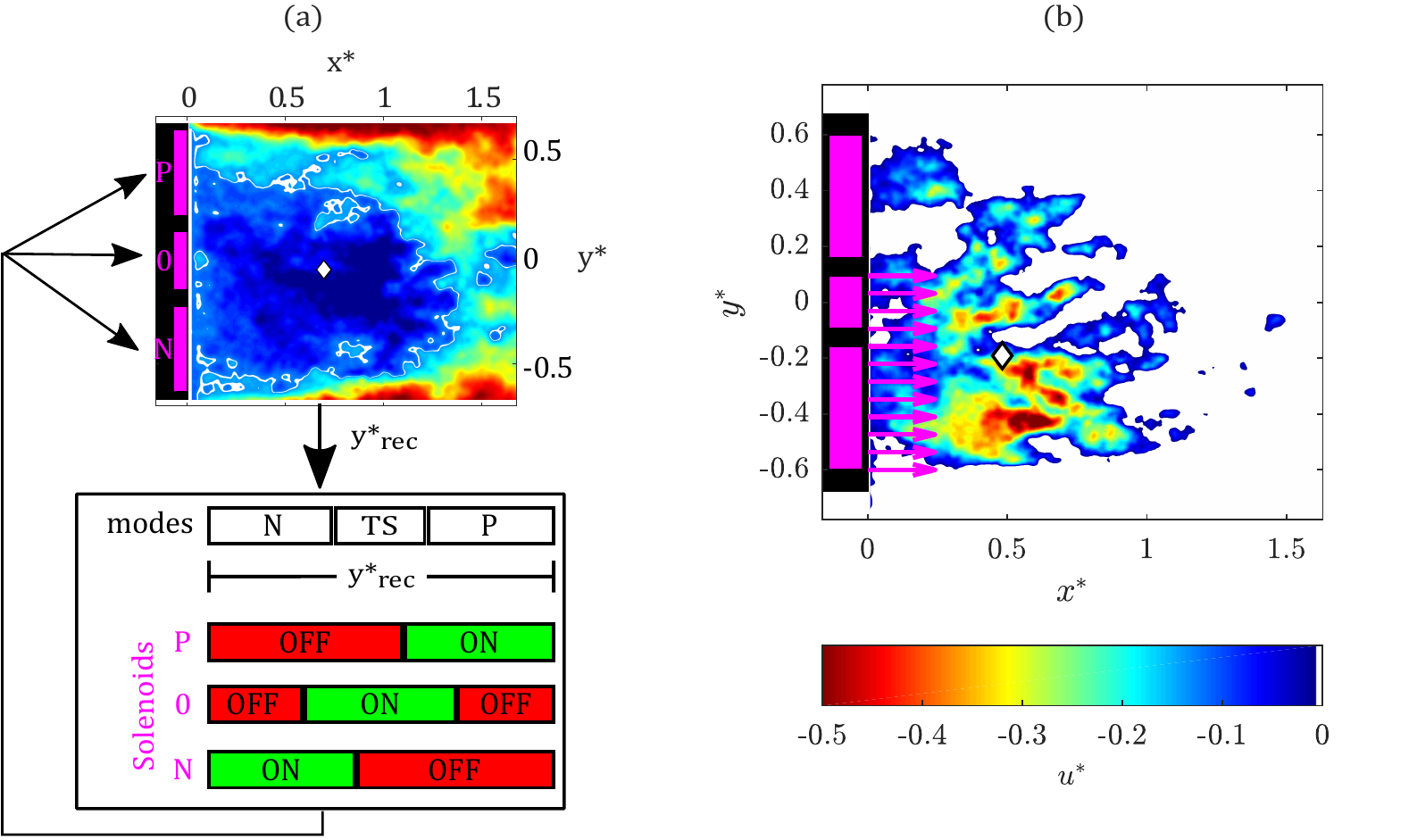}
	\caption{\label{fig:control_law_sketch}(a) Sketch of the opposition closed-loop control law, based on the detection of the spanwise position of the recirculation barycenter (white diamond), computed from the instantaneous velocity fields. (b) Illustration of the opposition control with the blowing regions chosen as a function of the position of the recirculation barycenter computed in real-time. Animated control law is available in Online Resource 2.}
	\end{figure*}

The principle of the closed-loop experiments is based on the computation in \emph{real-time} of the recirculation barycenter in the horizontal XY-plane at $z^*=1$, defining $y_{rec}^*$ as the control parameter. This tracking is then limited by the low acquisition frequency of the PIV setup ($f_{PIV}^*=2.8 \times 10^{-2}$). It means that we will act on the low-frequency large-scale dynamics of the wake. We then choose an opposition algorithm for the closed-loop experiments. It consists of blowing in the region where the recirculation barycenter is detected, as summarized in Fig.~\ref{fig:control_law_sketch}. This is consistent with the three modes defining the state of the wake: the two RSB modes and the TS mode. To these three modes correspond three blowing regions that are activated when the recirculation barycenter is detected in the related region. For each solenoids group we define a range for the spanwise location of the recirculation intensity barycenter $y_{rec}^*$ for which the group is activated. The instantaneous velocity fields and $y_{rec}^*$ are computed in \emph{real-time}. The control variable is updated with a period $T_{act}^*$, the minimum being $1/f_{PIV}^*$.

To make things more simple, we focus on the detection - blowing algorithm with all other actuation parameters fixed. The parameters for a closed-loop experiment are chosen based on the best or more interesting parameters found in the open-loop experiments. The actuation frequency $f_{j}^*$ and the mean jet velocity $\overbar{u_{jet}}$ are then chosen and set to fixed values before turning the control on.  

In the following, two types of actuation are used in the closed-loop experiments: a continuous blowing, denoted ``CL-CONT'' (``CL'' stands for closed-loop.), or a pulsed blowing with different actuation frequencies $f_{j}^*$. We will focus on two actuation frequencies: $f_{j}^*=0.2$ and $f_{j}^*=0.8$, respectively denoted ``CL-LF'' (for low frequency) and ``CL-HF'' (for high frequency). 

\subsection{Stabilization of the wake}

\subsubsection{Closed-loop with continuous blowing}

	\begin{figure*}
	\includegraphics{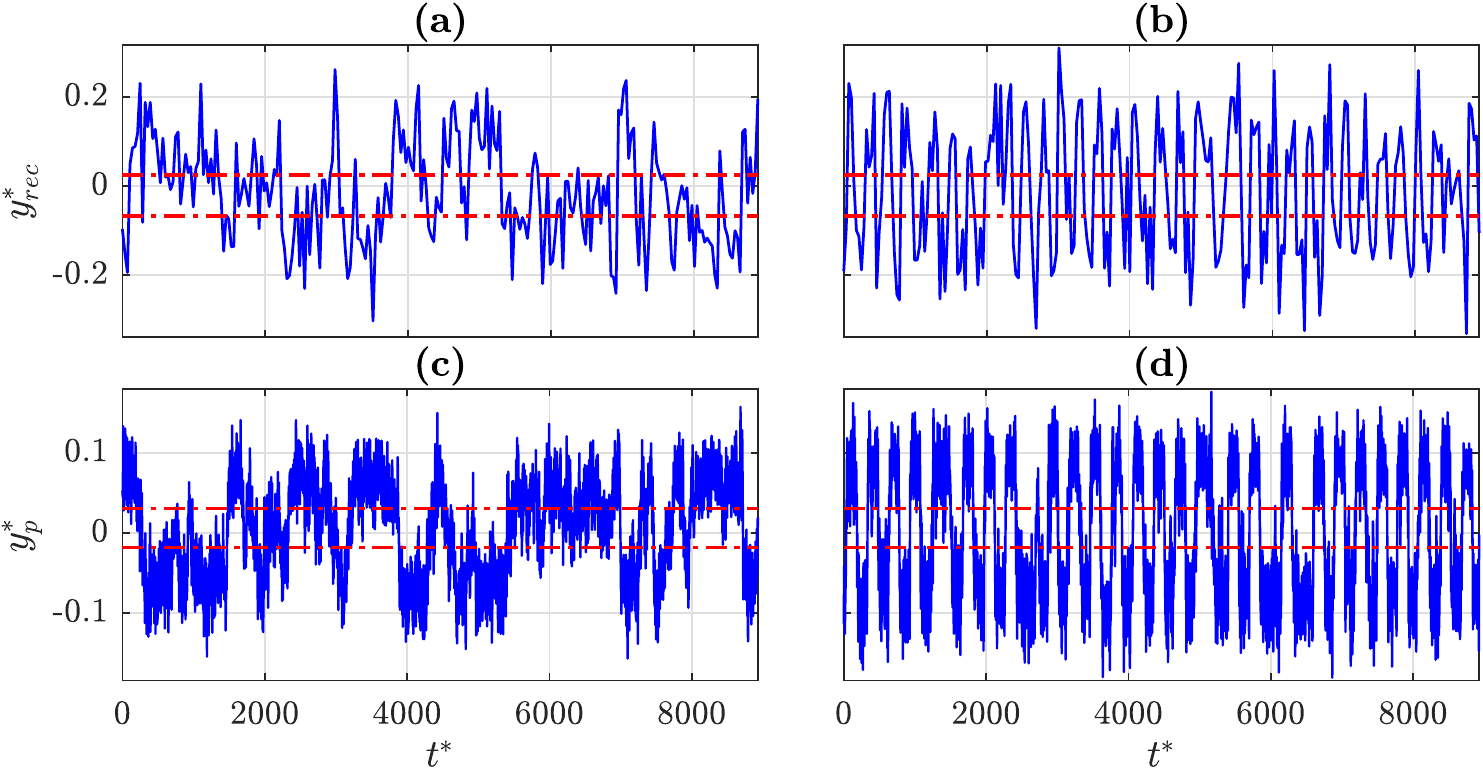}
	\caption{\label{fig:yP_yArec_ref_BF}Time series of the $y^*$-position of the recirculation barycenter and the pressure center for the reference case (a-c), compared to the closed-loop controlled case with continuous blowing jets (b-d). Thresholds are displayed in dash-dotted red.}
	\end{figure*}

The objective of these experiments is to control the wake dynamics which are characterized by the fluctuations of the spanwise location of the recirculation barycenter $y^*_{rec}(t)$. Typical time-series of the pressure and recirculation barycenters fluctuations for the reference case are displayed respectively in Fig.~\ref{fig:yP_yArec_ref_BF}(a) and Fig.~\ref{fig:yP_yArec_ref_BF}(c). When the closed-loop actuation CL-CONT is triggered, the dynamics of the barycenters [Figs.~\ref{fig:yP_yArec_ref_BF}(b) and (d)] are completely changed. Instead of random large and small fluctuations, the oscillations of the barycenters become very regular and periodic. The right-left switching is now imposed by the characteristic time scale of the closed-loop control law $T_{act}^*$. This observation holds also for the pulsed actuations.

	\begin{figure*}
	\includegraphics{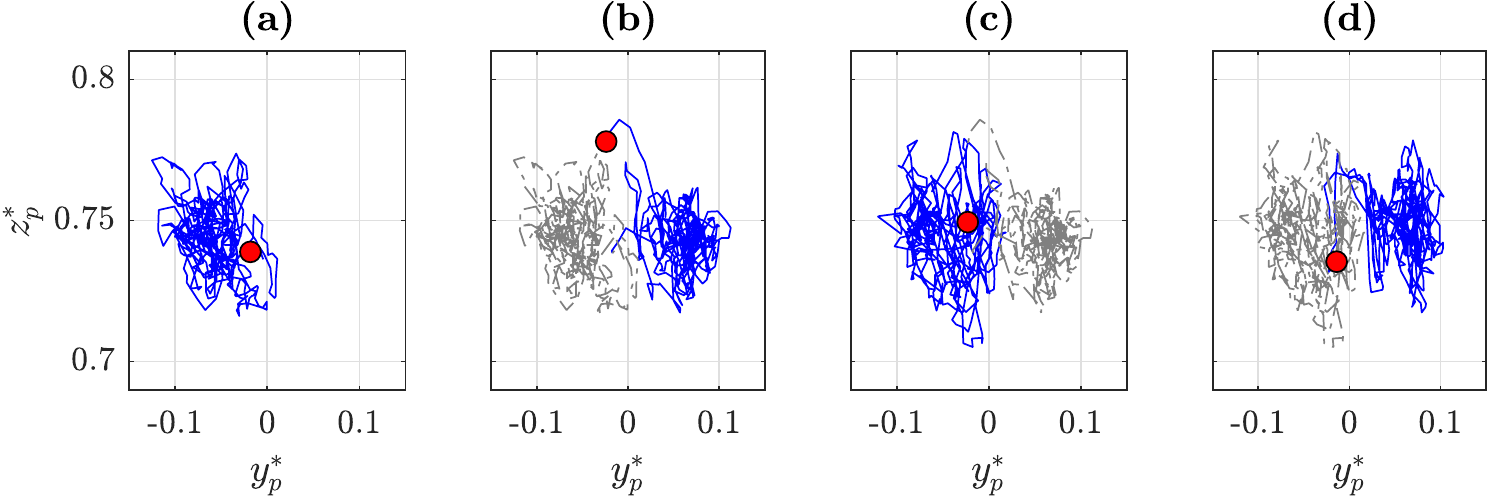}
	\caption{\label{fig:Gp_CL-CONT}Successive displacements of the pressure center $G_p$ for CL-CONT for (a) $\lbrack t_0^*:t_0^*+T_{act}^* \rbrack$, (b) $\lbrack t_0^*+T_{act}^*:t_0^*+2T_{act}^* \rbrack$, (c) $\lbrack t_0^*+2T_{act}^*:t_0^*+3T_{act}^* \rbrack$, and (d) $\lbrack t_0^*+3T_{act}^*:t_0^*+4T_{act}^* \rbrack$. The red circle is the last position in the corresponding sub-figure, the solid blue line shows the barycenter trajectory whereas the grey dashed line displays its previous one. Animated trajectory of $G_p$ for CL-CONT is available in Online Resource 3.}
	\end{figure*}

The trajectories of the pressure barycenter for CL-CONT during four successive control cycles are illustrated in Figs.~\ref{fig:Gp_CL-CONT}(a-d). It reveals that the flipping process occurs at the defined period $T_{act}^*$, only enforcing the presence of the two asymmetric modes, whereas the transient state becomes very short. The bimodal behavior is thus now completely driven by the opposition control law.

	\begin{figure*}
	\includegraphics[scale=1]{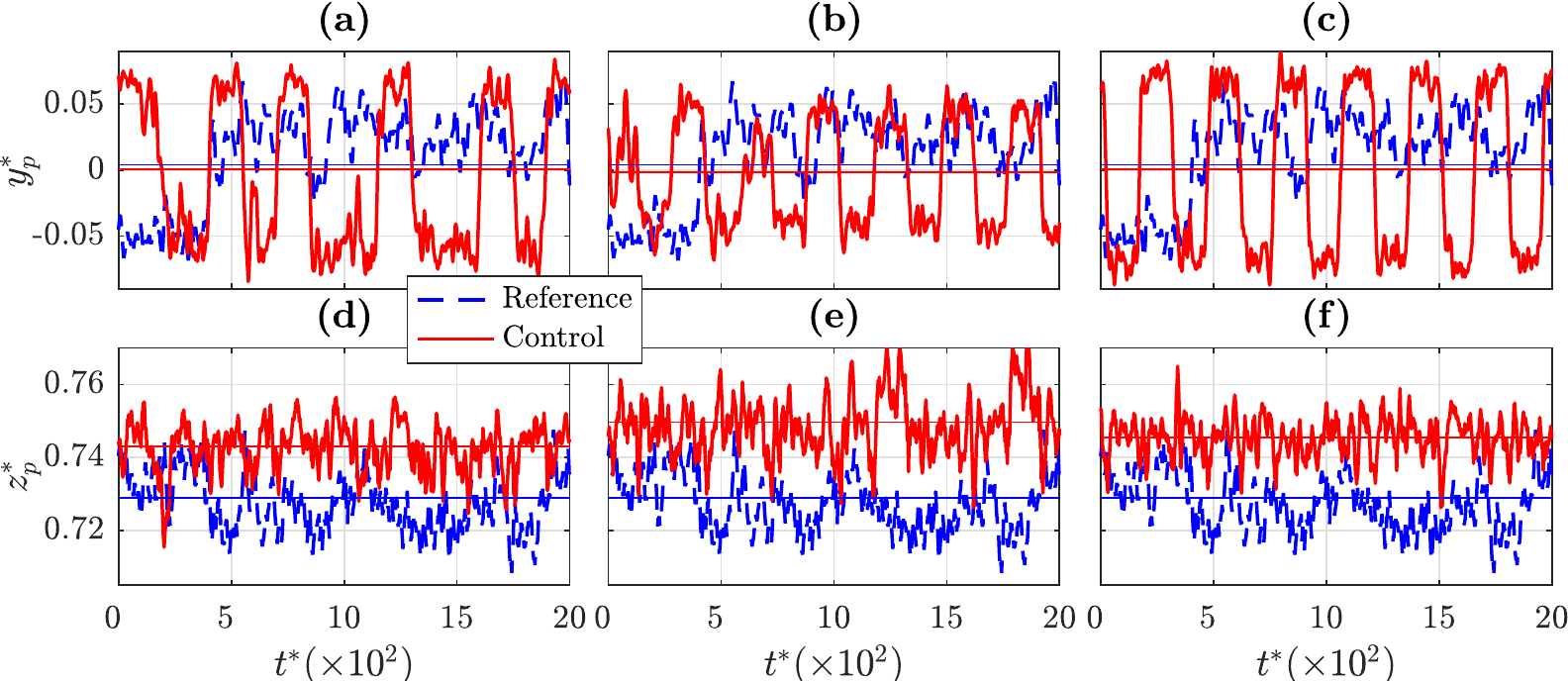}
	\caption{\label{fig:yPzP_ref_BF}Excerpts of $z_p^*$ and $z_p^*$ time series for (a-d) CL-CONT, (b-e) CL-LF, and (c-f) CL-HF (solid red), all compared to a reference time series (dashed blue). Data are smoothed over $\Delta t^*=15$ for clarity. The horizontal lines are the mean values.}
	\end{figure*}

\subsubsection{Closed-loop with pulsed blowing}

Introducing again pulsed jets instead of continuous blowing modifies the dynamics, depending on the jet frequency. On one hand, the dynamics obtained for CL-HF [Figs.~\ref{fig:yPzP_ref_BF}(c) and (f)] are very close to the ones for CL-CONT [Figs.~\ref{fig:yPzP_ref_BF}(a) and (d)]. On the other hand, Fig.~\ref{fig:yPzP_ref_BF}(b) shows for CL-LF slightly lower spanwise fluctuations, of the same order as the reference, whereas Fig.~\ref{fig:yPzP_ref_BF}(e) reveals a higher mean vertical position of the depression. This is probably due to the central jets which tend to symmetrize the wake and to attract the depression, as seen for the OL-0 case at this jet frequency.   

	\begin{figure*}
	\includegraphics[scale=1]{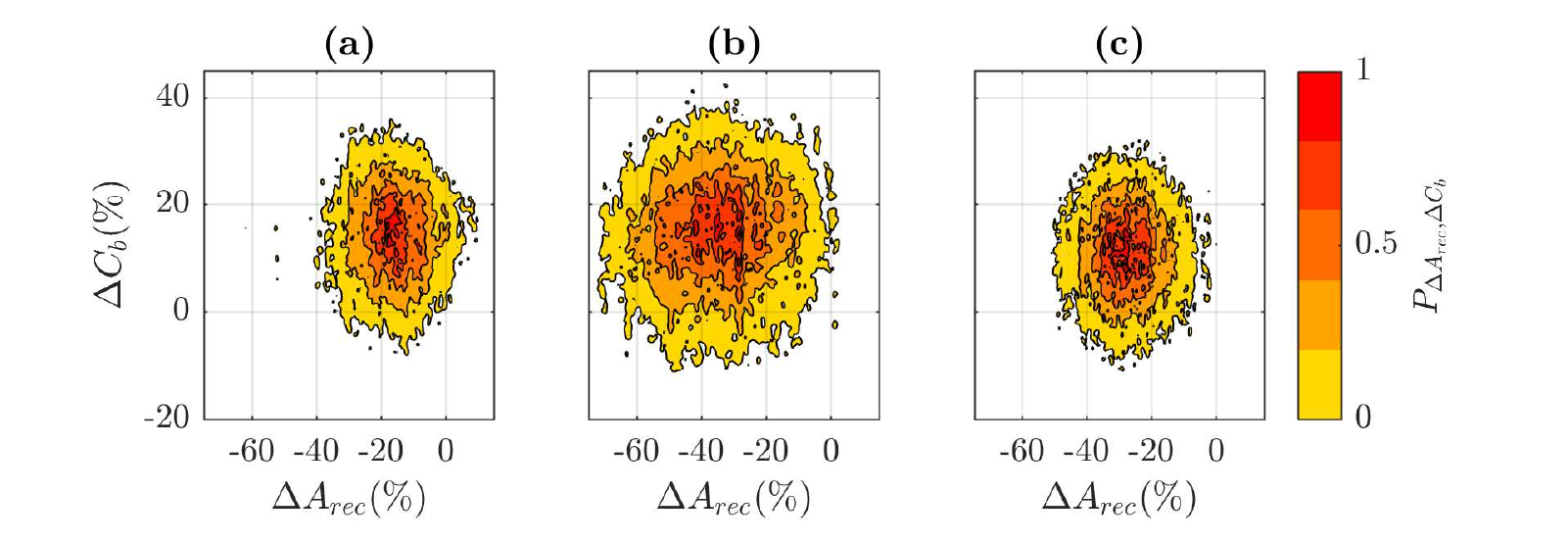}
	\caption{\label{fig:DeltaCb_DeltaArec_CL}$\Delta {C}_b$ with respect to $\Delta {A}_{rec}$ for (a) CL-CONT, (b) CL-LF, and (c) CL-HF.}
	\end{figure*}

It is then interesting to analyze how the pressure at the rear end of the model and the wake size are modified together, depending on the type of actuation used in the closed-loop experiments. Another state parameter for the wake is thus defined: the variation of the recirculation area $A_{rec}$ compared to the mean value of the reference case $A_{rec,ref}$ in the PIV-plane 
\begin{equation}
\Delta{A}_{rec}(t) = \frac{A_{rec}(t) - A_{rec,ref}}{A_{rec,ref}}.
\end{equation}
Instead of following the physical trajectories of the pressure barycenter over the rear part of the model, it is now possible to follow the trajectory of the wake in the state parameter ($\Delta{C}_{b}$, $\Delta{A}_{rec}$). Unfortunately, the acquisition frequency of the PIV fields is much smaller than the ones of the pressure fields, so that it is not possible to correlate all mean pressure coefficients to a corresponding recirculation area. Nevertheless, it is possible to correlate all PIV fields to a corresponding mean pressure coefficient and then plot the two state parameters. 

As shown in Figs.~\ref{fig:DeltaCb_DeltaArec_CL}, all closed-loop experiments lead to clear reduction of the recirculation area, i.e. of the size of the wake. The type of actuation has also a strong influence: pulsed jets are more efficient ($-40\%$ and $-32\%$ in average when $f_{j}^* =0.2$ and $f_{j}^* =0.8$ respectively) than continuous blowing ($-18\%$). On the contrary, the depression over the rear end increases in all cases, with the most important increase for $f_{j}^* =0.2$ ($+19\%$). The portion of the phase space explored by the wake is much larger for the CL-LF case [Fig.~\ref{fig:DeltaCb_DeltaArec_CL}(b)] than for the two other configurations but there is no clear relation between the changes in $\Delta{A}_{rec}$ and the ones in $\Delta{C}_{b}$. It also confirms the influence of the actuation frequency on the efficiency of the closed-loop control. The opposition algorithm has a different impact on the dynamics of the wake depending on the actuation frequency. It can be seen has a hint of the specific response of a chaotic system. Indeed, \citet{Grebogi1988} demonstrated that controlling a chaotic system may be more efficient if targeting the hidden frequency of the chaotic system, the UPO. If one can find the UPO of the system, then it should react to a forcing with small perturbations and the frequency of the UPO. The specific response of our system may be explained this way. Unfortunately, as stated previously, our time series were not long enough to evaluate the UPO of the system and verify this hypothesis.

	\begin{figure*}
	\includegraphics[scale=1]{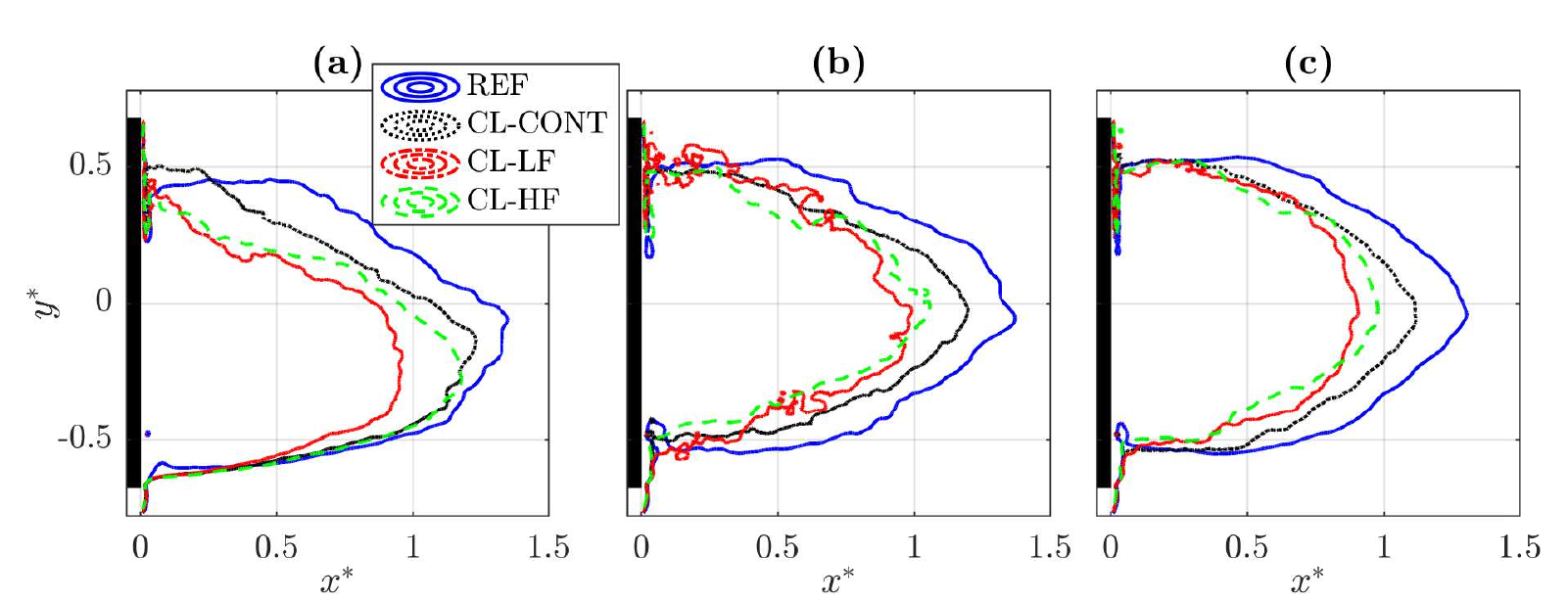}
	\caption{\label{fig:Arec_ref_BF}Contour of the recirculation area for the reference (blue solid), CL-CONT (black dotted), CL-LF (red dash-dotted) and CL-HF (green dashed) for (a) the N state, (b) the TS state and (c) the mean velocity field.}
	\end{figure*}

\subsubsection{Modification of the time-averaged wake}

Regarding the RSB modes in particular, the more the CL controls reduce the recirculation area, the more asymmetric the wake is, as highlighted in Fig.~\ref{fig:Arec_ref_BF}(a). Even if the TS state existence is reduced, the separatrices computed for this state [Fig.~\ref{fig:Arec_ref_BF}(b)] have the shape similar to the mean profiles [Fig.~\ref{fig:Arec_ref_BF}(c)]. This is due to the enhanced equiprobability of the BSR modes. It is noteworthy that the recirculation barycenter comes also closer to the bluff body for all types of actuations.

	\begin{figure*}
	\includegraphics[scale=1]{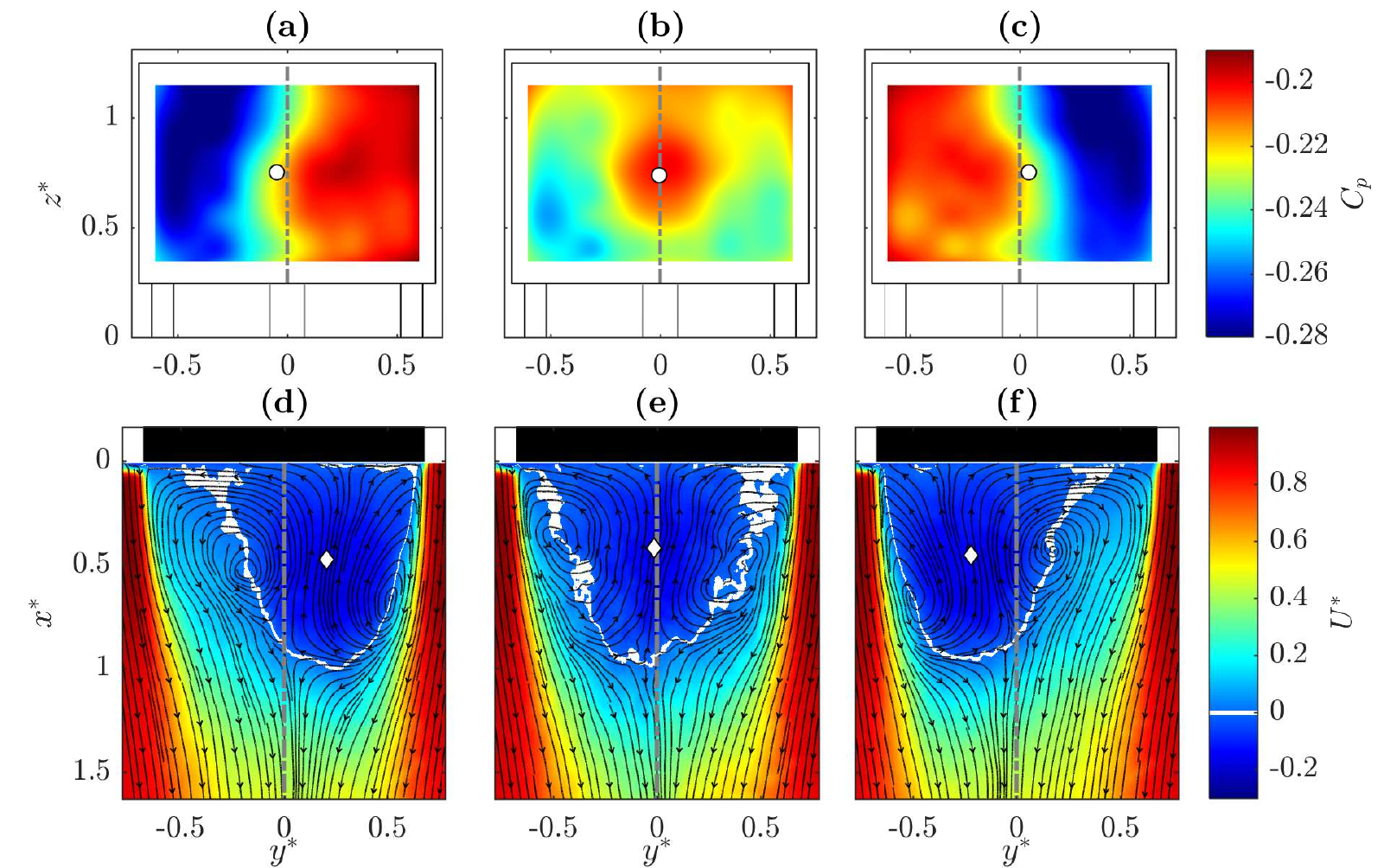}
	\caption{\label{fig:statesNTP_CL_LF}Conditional averages of the pressure coefficient at the rear body and the velocity field at $z^*=1$ for the modes (a-d) N, (b-e) TS and (c-f) P in the CL-LF case. The average position of the pressure center $G_p$ (white circle) and of the recirculation barycenter $G_{rec}$ (white diamond) are displayed. $y^*=0$ is also displayed (dash-dotted grey lines). Some streamlines are plotted over the streamwise component of the velocity to localize the coherent large-scale structures.}
	\end{figure*}

The conditional averages of the pressure and the velocity are computed and displayed in Fig.~\ref{fig:statesNTP_CL_LF} for CL-LF. The modes organization is affected by the forcing: the BSR modes appear more asymmetric and the TS state concentrates the higher pressures in the center. The mean rear pressure of the TS state is higher than the BSR one (-0.227 and -0.234).


\subsection{Creation of a multistable state}

\begin{figure*}
	\includegraphics[scale=1]{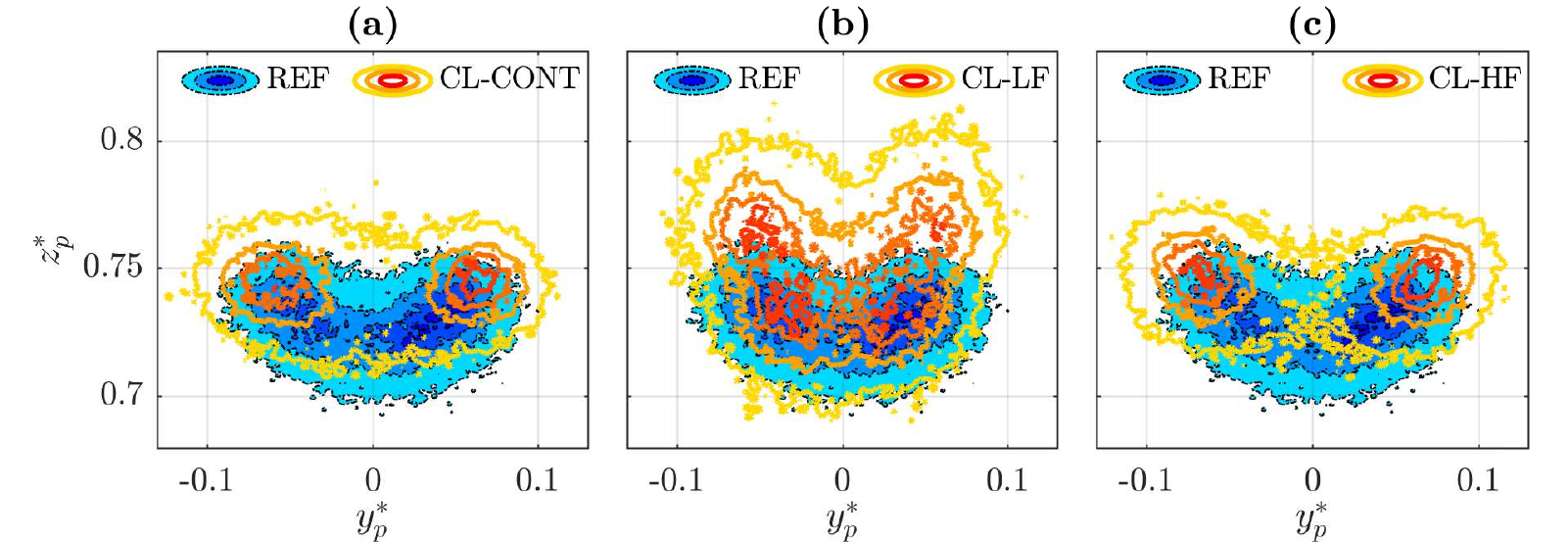}
	\caption{\label{fig:ZpYp_PDF_ref_CL}Joint PDF contour of ($y_p^*$,$z_p^*$) for the reference flow (blue) and various closed-loop controls (red): (a) CL-CONT, (b) CL-LF, and (c) CL-HF.}
	\end{figure*}

Another way to characterize the dynamics of the wake is to look again at the full trajectories of the pressure barycenter over the rear part of the model. For the three actuation types, the corresponding joint PDF are shown in Figs.~\ref{fig:ZpYp_PDF_ref_CL}, where the previously highlighted strange attractors are displayed in blue.

\begin{figure*}
	\includegraphics{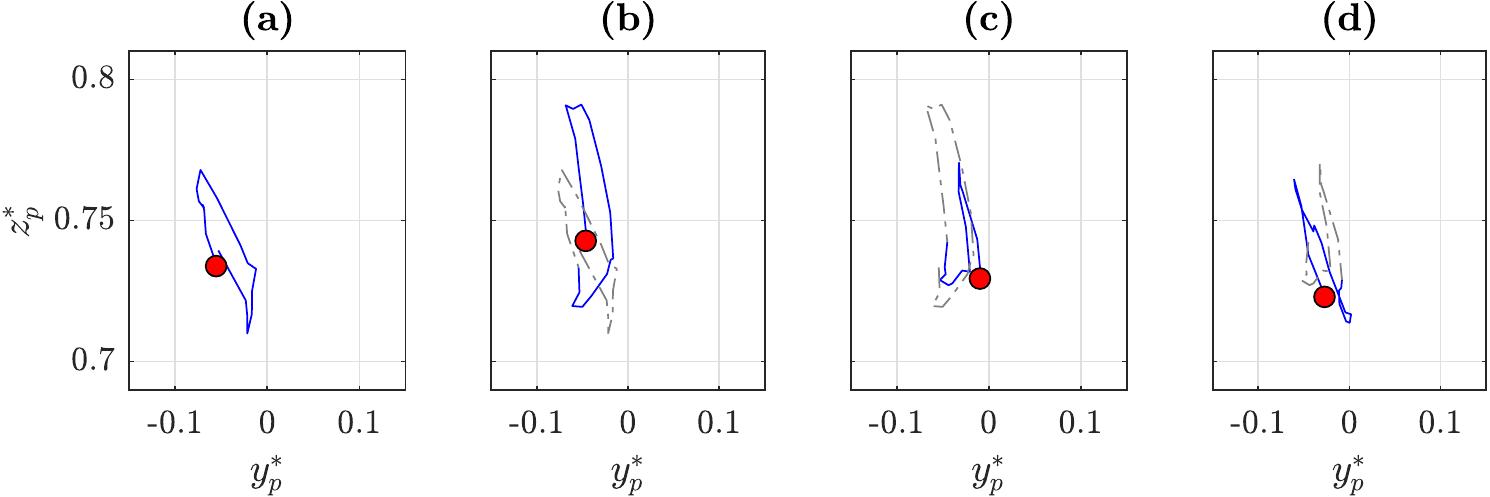}
	\caption{\label{fig:Gp_CL-LF}Successive displacements of the pressure center $G_p$ for CL-LF for (a) $\lbrack t_0^*:t_0^*+T_{j}^* \rbrack$, (b) $\lbrack t_0^*+T_{j}^*:t_0^*+2T_{j}^* \rbrack$, (c) $\lbrack t_0^*+2T_{j}^*:t_0^*+3T_{j}^* \rbrack$, and (d) $\lbrack t_0^*+3T_{j}^*:t_0^*+4T_{j}^* \rbrack$. The red circle is the last position in the corresponding sub-figure, the solid blue line shows the barycenter trajectory whereas the grey dashed line displays its previous one. Animated trajectory of $G_p$ for CL-LF is available in Online Resource 4.}
	\end{figure*}

When the recirculation barycenter is tracked and forced by the closed-loop algorithm with a continuous blowing, the PDF is also completely changed and exhibits two strong foci, as shown in Fig.~\ref{fig:ZpYp_PDF_ref_CL}(a). The transient mode is nearly suppressed and the wake switches quickly from one attractor to the other. This modification is also observed with CL-HF [Fig.~\ref{fig:ZpYp_PDF_ref_CL}(c)]. In both cases, the dynamics are marked by a decrease of $50 \%$ in the TS mode existence. Interestingly, the PDF is also strongly modified by the lower pulse frequency (CL-LF). In this case, one can see in Fig.~\ref{fig:ZpYp_PDF_ref_CL}(b) a butterfly shape with \emph{four} foci instead of two for all other cases. These results suggest the occurrence of multi-modal state with a right-left oscillation \emph{and} a top-bottom oscillation, which are undoubtedly enlarged due to the actuating frequency corresponding to the resonant frequency found in the spectral analysis of the $z_p^*(t)$ fluctuations [Fig.~\ref{fig:Sxx_yzP_ReH}(b)]. Figure \ref{fig:Gp_CL-LF} displays successive displacements of the pressure center for the CL-LF configuration. $z_p^*$ oscillates at the constant period $T_{j}^* = 1/f^*_{j}=5$ with a larger amplitude than the natural dynamics. 

Because of the sensitive dependence on the initial conditions, chaos was believed for a long time to be neither predictable nor controllable. The OGY control method proposed by Ott, Grebogi and Yorke \citep{Ott1990} changed completely this point of view. They showed that it is possible to use small time-dependent perturbations to force a chaotic attractors system in a time-periodic motion, defined from one of its UPOs \citep{Grebogi1988}. Since this pioneering work, control of chaotic systems has become a very active research field and many new control methods have been proposed such as linear-feedback \citep{Yassen2005}, adaptive control \citep{Tian2000}, impulsive or intermittent control \citep{Li2007}. Our result may be interpreted in this framework. The lower frequency may be the one associated with the hidden UPOs of the chaotic system. It may explain the specific effect observed with this type of actuation, in a way similar to the work on a turbulent jet \citep{Narayanan2013}. This harmonic resonance is also highlighted by \citet{Barros2016a}, through an antisymmetric open-loop forcing, close to the one resulting from our closed-loop control law. 

\section{Conclusions}
\label{conclusions}

Using global state parameters of the wake as the instantaneous barycenters of the spatially-averaged mean pressure coefficient or of the instantaneous recirculation area, it was possible to capture the dynamics of a 3D turbulent wake. It was also shown in a previous study that the dynamics of the wake, based on the fluctuations of pressure barycenter, are weakly chaotic. Having reduced the dynamics of the 3D wake to simple state parameters, it was then possible to design a simple closed-loop control experiment based on the real-time computations of the location of the recirculation area barycenter and a simple opposition algorithm. To our knowledge, this is also the first implementation of a closed-loop flow control experiment in a wind-tunnel facility based on the velocity fields computations, using \emph{real-time} PIV as a ``visual sensor".

The chaotic spanwise oscillations of the location of the recirculation (and pressure) barycenter is then forced into a well-controlled periodic oscillation at the time-scale imposed by the PIV setup and the algorithm. Moreover, a much larger reduction of the recirculation area is obtained when a pulsed actuation is used in the closed-loop control experiments, instead of a continuous blowing. When the frequency of the pulsed actuation used in the closed-loop experiments correspond to the natural shedding frequency the fluctuations in the phase space are strongly increased.

Finally, if the PDF of trajectories exhibits well-defined stable foci for most of the configurations, the low-frequency actuation case leads to a new four foci configuration, suggesting a multi-modal configuration with a right-left oscillation together with a top-bottom oscillation of the wake. These results show that once the turbulent wake dynamics has been reduced to a simple chaotic dynamics in physical space, it is indeed possible to design simple control laws to stabilize its dynamic around an unstable mode or to force its dynamics into simple harmonic oscillations. This work may find explanations in the framework of the theory of control of chaotic systems as proposed by \citet{Ott1990}. From a general point of view, it shows that there are new opportunities to investigate turbulent flows in the framework of dynamic and chaotic systems. It also opens the way to different control strategy of turbulent flows, as long as their dynamics can be reduced to a few state parameters. 
 
\begin{acknowledgements}
The measurements were performed in the Malavard wind tunnel (PRISME Institute, Orl\'eans, France) thanks to the involvement of the PRISME Institute team, especially P. Joseph and S. Loyer.
\end{acknowledgements}

\bibliographystyle{spbasic}      
\bibliography{Varon_ref}   

\begin{thebibliography}{84}
\providecommand{\natexlab}[1]{#1}
\providecommand{\url}[1]{{#1}}
\providecommand{\urlprefix}{URL }
\expandafter\ifx\csname urlstyle\endcsname\relax
  \providecommand{\doi}[1]{DOI~\discretionary{}{}{}#1}\else
  \providecommand{\doi}{DOI~\discretionary{}{}{}\begingroup
  \urlstyle{rm}\Url}\fi
\providecommand{\eprint}[2][]{\url{#2}}

\bibitem[{Ahmed et~al(1984)Ahmed, Ramm, and Faltin}]{Ahmed1984}
Ahmed SR, Ramm G, Faltin G (1984) Some salient features of the time-averaged
  ground vehicle wake. Tech. rep., SAE Technical Paper 840300,
  \doi{10.4271/840300}

\bibitem[{Auerbach et~al(1987)Auerbach, Cvitanovi\ifmmode~\acute{c}\else
  \'{c}\fi{}, Eckmann, Gunaratne, and Procaccia}]{Auerbach1987}
Auerbach D, Cvitanovi\ifmmode~\acute{c}\else \'{c}\fi{} P, Eckmann JP,
  Gunaratne G, Procaccia I (1987) Exploring chaotic motion through periodic
  orbits. Phys Rev Lett 58:2387--2389, \doi{10.1103/PhysRevLett.58.2387}

\bibitem[{Bailey et~al(2002)Bailey, Martinuzzi, and Kopp}]{Bailey2002}
Bailey SCC, Martinuzzi RJ, Kopp GA (2002) The effects of wall proximity on
  vortex shedding from a square cylinder: Three-dimensional effects. Phys
  Fluids 14(12):4160--4177, \doi{10.1063/1.1514972}

\bibitem[{Barros et~al(2016{\natexlab{a}})Barros, Bor\'{e}e, Noack, and
  Spohn}]{Barros2016a}
Barros D, Bor\'{e}e J, Noack BR, Spohn A (2016{\natexlab{a}}) Resonances in the
  forced turbulent wake past a 3d blunt body. Phys Fluids 28(6):065104,
  \doi{10.1063/1.4953176}

\bibitem[{Barros et~al(2016{\natexlab{b}})Barros, Bor\'{e}e, Noack, Spohn, and
  Ruiz}]{Barros2016b}
Barros D, Bor\'{e}e J, Noack BR, Spohn A, Ruiz T (2016{\natexlab{b}}) Bluff
  body drag manipulation using pulsed jets and coanda effect. J Fluid Mech
  805:422--459, \doi{10.1017/jfm.2016.508}

\bibitem[{Barros et~al(2017)Barros, Bor\'{e}e, Cadot, Spohn, and
  Noack}]{Barros2017}
Barros D, Bor\'{e}e J, Cadot O, Spohn A, Noack BR (2017) Forcing symmetry
  exchanges and flow reversals in turbulent wakes. J Fluid Mech 829,
  \doi{10.1017/jfm.2017.590}

\bibitem[{Beaudoin et~al(2004)Beaudoin, Cadot, Aider, Gosse, Parantho{\"e}n,
  Hamelin, Tissier, Allano, Mutabazi, Gonzales et~al}]{Beaudoin2004}
Beaudoin JF, Cadot O, Aider JL, Gosse K, Parantho{\"e}n P, Hamelin B, Tissier
  M, Allano D, Mutabazi I, Gonzales M, et~al (2004) Cavitation as a
  complementary tool for automotive aerodynamics. Exp Fluids 37(5):763--768,
  \doi{10.1007/s00348-004-0879-y}

\bibitem[{Brackston et~al(2016)Brackston, García de~la Cruz, Wynn, Rigas, and
  Morrison}]{Brackston2016}
Brackston RD, García de~la Cruz JM, Wynn A, Rigas G, Morrison JF (2016)
  Stochastic modeling and feedback control of bistability in a turbulent bluff
  body wake. J Fluid Mech 802:726--749, \doi{10.1017/jfm.2016.495}

\bibitem[{Broze and Hussain(1994)}]{Broze1994}
Broze G, Hussain F (1994) Nonlinear dynamics of forced transitional jets:
  {T}emporal attractors and transitions to chaos. In: Lin SP, Phillips WRC,
  Valentine DT (eds) Nonlinear Instability of Nonparallel Flows: IUTAM
  Symposium Potsdam, NY, USA July 26 -- 31, 1993, Springer, Berlin, Heidelberg,
  pp 459--473, \doi{10.1007/978-3-642-85084-4_38}

\bibitem[{Brunn and Nitsche(2006)}]{Brunn2006}
Brunn A, Nitsche W (2006) Active control of turbulent separated flows over
  slanted surfaces. Int J Heat Fluid Fl 27(5):748 -- 755,
  \doi{10.1016/j.ijheatfluidflow.2006.03.006}, special issue of the 6th
  International Symposium on Engineering Turbulence Modelling and Measurements
  – ETMM6

\bibitem[{Cadot et~al(2015)Cadot, Evrard, and Pastur}]{Cadot2015}
Cadot O, Evrard A, Pastur L (2015) Imperfect supercritical bifurcation in a
  three-dimensional turbulent wake. Phys Rev E 91(6):063005,
  \doi{10.1103/PhysRevE.91.063005}

\bibitem[{Cao(1997)}]{Cao1997}
Cao L (1997) Practical method for determining the minimum embedding dimension
  of a scalar time series. Phys D 110(1):43 -- 50,
  \doi{10.1016/S0167-2789(97)00118-8}

\bibitem[{Castro and Robins(1977)}]{Castro1977}
Castro IP, Robins AG (1977) The flow around a surface-mounted cube in uniform
  and turbulent streams. J Fluid Mech 79(2):307Ð335,
  \doi{10.1017/S0022112077000172}

\bibitem[{{Champagnat} et~al(2011){Champagnat}, {Plyer}, {Le Besnerais},
  {Leclaire}, {Davoust}, and {Le Saint}}]{Champagnat2011}
{Champagnat} F, {Plyer} A, {Le Besnerais} G, {Leclaire} B, {Davoust} S, {Le
  Saint} Y (2011) Fast and accurate {PIV} computation using highly parallel
  iterative correlation maximization. Exp Fluids 50:1169--1182,
  \doi{10.1007/s00348-011-1054-x}

\bibitem[{Corpetti et~al(2002)Corpetti, M{\'e}min, and Perez}]{Corpetti2002}
Corpetti T, M{\'e}min E, Perez P (2002) Dense estimation of fluid flows. IEEE T
  Pattern Anal 24(3):365--380, \doi{10.1109/34.990137}

\bibitem[{Corpetti et~al(2006)Corpetti, Heitz, Arroyo, M{\'e}min, and
  Santa-Cruz}]{Corpetti2006}
Corpetti T, Heitz D, Arroyo G, M{\'e}min E, Santa-Cruz A (2006) Fluid
  experimental flow estimation based on an optical-flow scheme. Exp Fluids
  40(1):80--97, \doi{10.1007/s00348-005-0048-y}

\bibitem[{Davoust et~al(2012)Davoust, Jacquin, and Leclaire}]{Davoust2012}
Davoust S, Jacquin L, Leclaire B (2012) Dynamics of m=0 and m=1 modes and of
  streamwise votrices in a turbulent axisymmetric mixing layer. J Fluid Mech
  709:408--444, \doi{10.1017/jfm.2012.342}

\bibitem[{Douay et~al(2013)Douay, Faure, and Lusseyran}]{Douay2013}
Douay CL, Faure TM, Lusseyran F (2013) Stereoscopic {PIV} using optical flow:
  application to a cavity recirculation. Exp Fluids 54(8):1579,
  \doi{10.1007/s00348-013-1579-2}

\bibitem[{Duell and George(1999)}]{Duell1999}
Duell EG, George AR (1999) Experimental study of a ground vehicle body unsteady
  near wake. In: SAE Technical Paper, SAE International,
  \doi{10.4271/1999-01-0812}

\bibitem[{Eckmann et~al(1986)Eckmann, Kamphorst, Ruelle, and
  Ciliberto}]{Eckmann1986}
Eckmann JP, Kamphorst SO, Ruelle D, Ciliberto S (1986) Liapunov exponents from
  time series. Phys Rev A 34:4971--4979, \doi{10.1103/PhysRevA.34.4971}

\bibitem[{{Eulalie}(2014)}]{Eulalie2014phd}
{Eulalie} Y (2014) {Aerodynamic analysis and drag reduction around an Ahmed
  bluff body}. PhD thesis, Universit{\'e} de Bordeaux, Bordeaux, France

\bibitem[{Evrard et~al(2016)Evrard, Cadot, Herbert, Ricot, Vigneron, and
  D\'{e}lery}]{Evrard2016}
Evrard A, Cadot O, Herbert V, Ricot D, Vigneron R, D\'{e}lery J (2016) Fluid
  force and symmetry breaking modes of a 3d bluff body with a base cavity. J
  Fluid Mech 61:99 -- 114, \doi{10.1016/j.jfluidstructs.2015.12.001}

\bibitem[{Evstafyeva et~al(2017)Evstafyeva, Morgans, and
  Dalla~Longa}]{Evstafyeva2017}
Evstafyeva O, Morgans AS, Dalla~Longa L (2017) {Simulation and feedback control
  of the Ahmed body flow exhibiting symmetry breaking behaviour}. J Fluid Mech
  817, \doi{10.1017/jfm.2017.118}

\bibitem[{Faranda et~al(2017)Faranda, Sato, Saint-Michel, Wiertel, Padilla,
  Dubrulle, and Daviaud}]{Faranda2017}
Faranda D, Sato Y, Saint-Michel B, Wiertel C, Padilla V, Dubrulle B, Daviaud F
  (2017) Stochastic chaos in a turbulent swirling flow. Phys Rev Lett
  119(1):014502, \doi{10.1103/PhysRevLett.119.014502}

\bibitem[{Faure et~al(2010)Faure, Douay, Mochki, Lusseyrand, and
  Quenot}]{Faure2010}
Faure TM, Douay CL, Mochki S, Lusseyrand F, Quenot G (2010) {Stereoscopic PIV
  using optical flow: Investigation of a recirculating cavity flow}. In: 8th
  International ERCOFTAC Symposium on Engineering Turbulence Modeling and
  Measurements, Marseille : France, vol~3, pp 905--910

\bibitem[{Gautier and Aider(2014)}]{Gautier2014}
Gautier N, Aider JL (2014) Feed-forward control of a perturbed backward-facing
  step flow. J Fluid Mech 759:181--196, \doi{10.1017/jfm.2014.518}

\bibitem[{Gautier and Aider(2015)}]{Gautier2015a}
Gautier N, Aider JL (2015) Frequency-lock reactive control of a separated flow
  enabled by visual sensors. Exp Fluids 56(1):1--10,
  \doi{10.1007/s00348-014-1869-3}

\bibitem[{{Gautier} and {Aider}(2015)}]{Gautier2015c}
{Gautier} N, {Aider} JL (2015) Real-time planar flow velocity measurements
  using an optical flow algorithm implemented on {GPU}. J Vis 18:277--286,
  \doi{10.1007/s12650-014-0222-5}

\bibitem[{Gautier et~al(2015)Gautier, Aider, Duriez, Noack, Segond, and
  Abel}]{Gautier2015b}
Gautier N, Aider JL, Duriez T, Noack BR, Segond M, Abel M (2015) Closed-loop
  separation control using machine learning. J Fluid Mech 770:442--457,
  \doi{10.1017/jfm.2015.95}

\bibitem[{Gentile et~al(2016)Gentile, Schrijer, Van~Oudheusden, and
  Scarano}]{Gentile2016}
Gentile V, Schrijer FFJ, Van~Oudheusden BW, Scarano F (2016) Low-frequency
  behavior of the turbulent axisymmetric near-wake. Phys Fluids 28(6):065102,
  \doi{10.1063/1.4953150}, \eprint{http://dx.doi.org/10.1063/1.4953150}

\bibitem[{Grandemange et~al(2012)Grandemange, Cadot, and
  Gohlke}]{Grandemange2012a}
Grandemange M, Cadot O, Gohlke M (2012) Reflectional symmetry breaking of the
  separated flow over three-dimensional bluff bodies. Phys Rev E 86(3):035302,
  \doi{10.1103/PhysRevE.86.035302}

\bibitem[{Grandemange et~al(2013)Grandemange, Gohlke, and
  Cadot}]{Grandemange2013c}
Grandemange M, Gohlke M, Cadot O (2013) Bi-stability in the turbulent wake past
  parallelepiped bodies with various aspect ratios and wall effects. Phys
  Fluids 25(9):095103, \doi{10.1063/1.4820372}

\bibitem[{{Grandemange} et~al(2013){Grandemange}, {Gohlke}, and
  {Cadot}}]{Grandemange2013a}
{Grandemange} M, {Gohlke} M, {Cadot} O (2013) Turbulent wake past a
  three-dimensional blunt body, part 1: Global modes and bi-stability. J Fluid
  Mech 722:51--84, \doi{10.1017/jfm.2013.83}

\bibitem[{Grandemange et~al(2014{\natexlab{a}})Grandemange, Gohlke, and
  Cadot}]{Grandemange2014b}
Grandemange M, Gohlke M, Cadot O (2014{\natexlab{a}}) Statistical axisymmetry
  of the turbulent sphere wake. Exp Fluids 55(11):1838,
  \doi{10.1007/s00348-014-1838-x}

\bibitem[{Grandemange et~al(2014{\natexlab{b}})Grandemange, Gohlke, and
  Cadot}]{Grandemange2014a}
Grandemange M, Gohlke M, Cadot O (2014{\natexlab{b}}) Turbulent wake past a
  three-dimensional blunt body, part 2: {E}xperimental sensitivity analysis. J
  Fluid Mech 752:439 -- 461, \doi{10.1017/jfm.2014.345}

\bibitem[{Grebogi et~al(1988)Grebogi, Ott, and Yorke}]{Grebogi1988}
Grebogi C, Ott E, Yorke JA (1988) Unstable periodic orbits and the dimensions
  of multifractal chaotic attractors. Phys Rev A 37:1711--1724,
  \doi{10.1103/PhysRevA.37.1711}

\bibitem[{Gumowski et~al(2008)Gumowski, Miedzik, {Goujon-Durand}, Jenffer, and
  Wesfreid}]{Gumowski2008}
Gumowski K, Miedzik J, {Goujon-Durand} S, Jenffer P, Wesfreid J (2008)
  Transition to a time-dependent state of fluid flow in the wake of a sphere.
  Phys Rev E 77(5):055308, \doi{10.1103/PhysRevE.77.055308}

\bibitem[{Hauet et~al(2008)Hauet, Kruger, Krajewski, Bradley, Muste, Creutin,
  and Wilson}]{Hauet2008}
Hauet A, Kruger A, Krajewski WF, Bradley A, Muste M, Creutin JD, Wilson M
  (2008) Experimental system for real-time discharge estimation using an
  image-based method. J Hydrol Eng 13(2):105--110,
  \doi{10.1061/(ASCE)1084-0699(2008)13:2(105)}

\bibitem[{Hinterberger et~al(2004)Hinterberger, Garc{\'i}a-Villalba, and
  Rodi}]{Hinterberger2004}
Hinterberger C, Garc{\'i}a-Villalba M, Rodi W (2004) Large eddy simulation of
  flow around the {A}hmed body. In: McCallen R, Browand F, Ross J (eds) The
  Aerodynamics of Heavy Vehicles: {T}rucks, Buses, and Trains, Springer,
  Berlin, Heidelberg, pp 77--87, \doi{10.1007/978-3-540-44419-0_7}

\bibitem[{Hucho(2013)}]{Hucho2013}
Hucho WH (2013) Aerodynamics of road vehicles: {F}rom fluid mechanics to
  vehicle engineering. Elsevier

\bibitem[{Iriarte~Munoz et~al(2009)Iriarte~Munoz, Dellavale, Sonnaillon, and
  Bonetto}]{IriarteMunoz2009}
Iriarte~Munoz JM, Dellavale D, Sonnaillon MO, Bonetto FJ (2009) {Real-time
  particle image velocimetry based on FPGA technology}. In: 2009 5th Southern
  Conference on Programmable Logic (SPL), pp 147--152,
  \doi{10.1109/SPL.2009.4914899}

\bibitem[{Joseph et~al(2012)Joseph, Amandol{\`e}se, and Aider}]{Joseph2012}
Joseph P, Amandol{\`e}se X, Aider JL (2012) Drag reduction on the 25$^\circ$
  slant angle ahmed reference body using pulsed jets. Exp Fluids
  52(5):1169--1185, \doi{10.1007/s00348-011-1245-5}

\bibitem[{Kaiser et~al(2014)Kaiser, Noack, Cordier, Spohn, Segond, Abel,
  Daviller, Östh, Krajnovi\`{c}, Niven, and et~al.}]{Kaiser2014}
Kaiser E, Noack BR, Cordier L, Spohn A, Segond M, Abel M, Daviller G, Östh J,
  Krajnovi\`{c} S, Niven RK, et~al (2014) Cluster-based reduced-order modelling
  of a mixing layer. J Fluid Mech 754:365--414, \doi{10.1017/jfm.2014.355}

\bibitem[{Khalighi et~al(2012)Khalighi, Chen, and Iaccarino}]{Khalighi2012}
Khalighi B, Chen K, Iaccarino GG (2012) Unsteady aerodynamic flow investigation
  around a simplified square-back road vehicle with drag reduction devices. J
  Fluids Eng 134(6):061101, \doi{10.1115/1.4006643}

\bibitem[{Klotz et~al(2014)Klotz, Goujon-Durand, Rokicki, and
  Wesfreid}]{Klotz2014}
Klotz L, Goujon-Durand S, Rokicki J, Wesfreid J (2014) Experimental
  investigation of flow behind a cube for moderate reynolds numbers. J Fluid
  Mech 750:73 -- 98, \doi{10.1017/jfm.2014.236}

\bibitem[{Krajnovi\'{c} and Davidson(2003)}]{Krajnovic2003}
Krajnovi\'{c} S, Davidson L (2003) Numerical study of the flow around a
  bus-shaped body. J Fluid Eng 125(3):500--509, \doi{10.1115/1.1567305}

\bibitem[{Krajnovi\'{c} and Davidson(2005)}]{Krajnovic2005}
Krajnovi\'{c} S, Davidson L (2005) Flow around a simplified car, part 1:
  {L}arge eddy simulation. J Fluid Eng 127(5):907--918, \doi{10.1115/1.1989371}

\bibitem[{Lahaye et~al(2014)Lahaye, Leroy, and Kourta}]{Lahaye2014}
Lahaye A, Leroy A, Kourta A (2014) Aerodynamic characterisation of a
  square-back bluff body flow. Int J of Aerodynamics 4(1-2):43--60,
  \doi{10.1504/IJAD.2014.057804}

\bibitem[{{Le Besnerais} and Champagnat(2005)}]{LeBesnerais2005}
{Le Besnerais} G, Champagnat F (2005) Dense optical flow by iterative local
  window registration. In: Image Processing. ICIP. IEEE International
  Conference, IEEE, vol~1, pp 137--140, \doi{10.1109/ICIP.2005.1529706}

\bibitem[{Li et~al(2007)Li, Liao, and Huang}]{Li2007}
Li C, Liao X, Huang T (2007) Exponential stabilization of chaotic systems with
  delay by periodically intermittent control. Chaos 17(1):013103,
  \doi{10.1063/1.2430394}

\bibitem[{Li et~al(2016)Li, Barros, Bor{\'e}e, Cadot, Noack, and
  Cordier}]{Li2016}
Li R, Barros D, Bor{\'e}e J, Cadot O, Noack BR, Cordier L (2016) Feedback
  control of bimodal wake dynamics. Exp Fluids 57(10):158,
  \doi{10.1007/s00348-016-2245-2}

\bibitem[{Lienhart et~al(2002)Lienhart, Stoots, and Becker}]{Lienhart2002}
Lienhart H, Stoots C, Becker S (2002) Flow and turbulence structures in the
  wake of a simplified car model ({A}hmed modell). In: Wagner S, Rist U,
  Heinemann HJ, Hilbig R (eds) New Results in Numerical and Experimental Fluid
  Mechanics III: Contributions to the 12th STAB/DGLR Symposium Stuttgart,
  Germany 2000, Springer, Berlin, Heidelberg, pp 323--330,
  \doi{10.1007/978-3-540-45466-3_39}

\bibitem[{Ma et~al(2013)Ma, Lin, and Lai}]{Ma2013}
Ma H, Lin W, Lai YC (2013) Detecting unstable periodic orbits in
  high-dimensional chaotic systems from time series: {R}econstruction meeting
  with adaptation. Phys Rev E 87:050901, \doi{10.1103/PhysRevE.87.050901}

\bibitem[{Mandelbrot(1982)}]{Mandelbrot1982}
Mandelbrot B (1982) The fractal geometry of nature. Henry Holt and Company,
  New-York

\bibitem[{McArthur et~al(2016)McArthur, Burton, Thompson, and
  Sheridan}]{McArthur2016}
McArthur D, Burton D, Thompson M, Sheridan J (2016) On the near wake of a
  simplified heavy vehicle. J Fluid Struct 66:293 -- 314,
  \doi{10.1016/j.jfluidstructs.2016.07.011}

\bibitem[{Morel(1978)}]{Morel1978}
Morel T (1978) The effect of base slant on the flow pattern and drag of
  three-dimensional bodies with blunt ends. In: Sovran G, Morel T, Mason WT
  (eds) Aerodynamic Drag Mechanisms of Bluff Bodies and Road Vehicles, Springer
  US, Boston, MA, pp 191--226, \doi{10.1007/978-1-4684-8434-2_8}

\bibitem[{Narayanan et~al(2013)Narayanan, Gunaratne, and
  Hussain}]{Narayanan2013}
Narayanan S, Gunaratne GH, Hussain F (2013) A dynamical systems approach to the
  control of chaotic dynamics in a spatiotemporal jet flow. Chaos 23(3):033133,
  \doi{10.1063/1.4820819}, \eprint{http://dx.doi.org/10.1063/1.4820819}

\bibitem[{NVIDIA(2010)}]{NVIDIAFERMI}
NVIDIA (2010) Whitepaper. NVIDIA's Next Generation CUDA Compute Architecture:
  Fermi

\bibitem[{\"Osth et~al(2014)\"Osth, Noack, Krajnovi\'{c}, Barros, and
  Bor\'{e}e}]{Oesth2014}
\"Osth J, Noack BR, Krajnovi\'{c} S, Barros D, Bor\'{e}e J (2014) On the need
  for a nonlinear subscale turbulence term in pod models as exemplified for a
  high-reynolds-number flow over an ahmed body. J Fluid Mech 747:518--544,
  \doi{10.1017/jfm.2014.168}

\bibitem[{Ott et~al(1990)Ott, Grebogi, and Yorke}]{Ott1990}
Ott E, Grebogi C, Yorke JA (1990) Controlling chaos. Phys Rev Lett
  64(11):1196--1199, \doi{10.1103/PhysRevLett.64.1196}

\bibitem[{Pan et~al(2015)Pan, Xue, Xu, Wang, and Wei}]{Pan2015}
Pan C, Xue D, Xu Y, Wang J, Wei R (2015) {Evaluating the accuracy performance
  of Lucas-Kanade algorithm in the circumstance of PIV application}. Sci China
  Phys Mech 58:104704, \doi{10.1007/s11433-015-5719-y}

\bibitem[{Parezanovi\'{c} and Cadot(2012)}]{Parezanovic2012}
Parezanovi\'{c} V, Cadot O (2012) Experimental sensitivity analysis of the
  global properties of a two-dimensional turbulent wake. J Fluid Mech
  693:115--149, \doi{10.1017/jfm.2011.495}

\bibitem[{Pastoor et~al(2008)Pastoor, Henning, Noack, King, and
  Tadmor}]{Pastoor2008}
Pastoor M, Henning L, Noack BR, King R, Tadmor G (2008) Feedback shear layer
  control for bluff body drag reduction. J Fluid Mech 608:161–196,
  \doi{10.1017/S0022112008002073}

\bibitem[{Pavia et~al(2017)Pavia, Passmore, and Sardu}]{Pavia2017}
Pavia G, Passmore M, Sardu C (2017) Evolution of the bi-stable wake of a
  square-back automotive shape. Exp Fluids 59(1),
  \doi{10.1007/s00348-017-2473-0}

\bibitem[{Plyer et~al(2016)Plyer, Le~Besnerais, and Champagnat}]{Plyer2016}
Plyer A, Le~Besnerais G, Champagnat F (2016) Massively parallel lucas kanade
  optical flow for real-time video processing applications. J Real-Time Image
  Proc 11(4):713--730, \doi{10.1007/s11554-014-0423-0}

\bibitem[{Provenzale et~al(1992)Provenzale, Smith, Vio, and
  Murante}]{Provenzale1992}
Provenzale A, Smith L, Vio R, Murante G (1992) Distinguishing between
  low-dimensional dynamics and randomness in measured time series. Physica D
  58(1):31 -- 49, \doi{10.1016/0167-2789(92)90100-2}

\bibitem[{Qu{\'e}not et~al(1998)Qu{\'e}not, Pakleza, and
  Kowalewski}]{Quenot1998}
Qu{\'e}not GM, Pakleza J, Kowalewski TA (1998) Particle image velocimetry with
  optical flow. Exp Fluids 25(3):177--189, \doi{10.1007/s003480050222}

\bibitem[{Rao et~al(2018)Rao, Minelli, Basara, and Krajnovi\'{c}}]{Rao2018}
Rao A, Minelli G, Basara B, Krajnovi\'{c} S (2018) Investigation of the
  near-wake flow topology of a simplified heavy vehicle using {PANS}
  simulations. J Wind Eng Ind Aerod 183:243--272,
  \doi{10.1016/j.jweia.2018.09.019}

\bibitem[{Roberts(2012)}]{Roberts2012phd}
Roberts JW (2012) Control of underactuated fluid-body systems with real-time
  particle image velocimetry. PhD thesis, MIT, Boston, USA

\bibitem[{Sartor et~al(2012)Sartor, Losfeld, and Bur}]{Sartor2012}
Sartor F, Losfeld G, Bur R (2012) {PIV} study on a shock-induced transition in
  transonic flow. Exp Fluids 53:815--827, \doi{10.1007/s00348-012-1330-4}

\bibitem[{Siegel et~al(2003)Siegel, Cohen, {McLaughlin}, and
  Myatt}]{Siegel2003}
Siegel S, Cohen K, {McLaughlin} T, Myatt J (2003) Real-time particle image
  velocimetry for closed-loop flow control studies. In: 41st Aerospace Sciences
  Meeting and Exhibit, Aerospace Sciences Meetings, Reno, USA,
  \doi{10.2514/6.2003-920}

\bibitem[{Sigurdson(1995)}]{Sigurdson1995}
Sigurdson LW (1995) The structure and control of a turbulent reattaching flow.
  J Fluid Mech 298:139 -- 165, \doi{10.1017/S0022112095003259}

\bibitem[{Stanislas et~al(2008)Stanislas, Okamoto, K{\"a}hler, Westerweel, and
  Scarano}]{Stanislas2008}
Stanislas M, Okamoto K, K{\"a}hler CJ, Westerweel J, Scarano F (2008) {Main
  results of the third international PIV Challenge}. Exp Fluids 45(1):27--71,
  \doi{10.1007/s00348-008-0462-z}

\bibitem[{{Takens}(1981)}]{Takens1981}
{Takens} F (1981) Detecting strange attractors in turbulence, Springer Berlin
  Heidelberg, Berlin, Heidelberg, pp 366--381. \doi{10.1007/BFb0091924}

\bibitem[{Tian and Yu(2000)}]{Tian2000}
Tian YP, Yu X (2000) Adaptive control of chaotic dynamical systems using
  invariant manifold approach. IEEE T Circuits-I 47(10):1537--1542,
  \doi{10.1109/81.886986}

\bibitem[{Varon et~al(2017)Varon, Eulalie, Edwige, Gilotte, and
  Aider}]{Varon2017}
Varon E, Eulalie Y, Edwige S, Gilotte P, Aider JL (2017) Chaotic dynamics of
  large-scale structures in a turbulent wake. Phys Rev Fluids 2:034604,
  \doi{10.1103/PhysRevFluids.2.034604}

\bibitem[{Venning et~al(2017)Venning, {Lo Jacono}, Burton, Thompson, and
  Sheridan}]{Venning2017}
Venning J, {Lo Jacono} D, Burton D, Thompson MC, Sheridan J (2017) {The nature
  of the vortical structures in the near wake of the Ahmed body}. In: Proc.
  I.Mech.E. Part D: J. Automobile Engineering, \doi{10.1177/0954407017690683}

\bibitem[{Volpe et~al(2015)Volpe, Devinant, and Kourta}]{Volpe2015}
Volpe R, Devinant P, Kourta A (2015) {Experimental characterization of the
  unsteady natural wake of the full-scale square back Ahmed body: Flow
  bi-stability and spectral analysis}. Exp Fluids 56(5):99,
  \doi{10.1007/s00348-015-1972-0}

\bibitem[{Willert et~al(2010)Willert, Munson, and Gharib}]{Willert2010}
Willert C, Munson M, Gharib M (2010) {Real-time particle image velocimetry for
  closed-loop flow control applications}. In: 15th Internation Symposium on
  Applications of Laser Techniques to Fluid Mechanics, Lisbon, Portugal

\bibitem[{Williamson(1996)}]{Williamson1996}
Williamson CHK (1996) Vortex dynamics in the cylinder wake. Annu Rev Fluid Mech
  28(1):477--539, \doi{10.1146/annurev.fl.28.010196.002401}

\bibitem[{Yang et~al(2015)Yang, Liu, Wu, Liu, and Zhang}]{Yang2015}
Yang J, Liu M, Wu G, Liu Q, Zhang X (2015) Low-frequency characteristics in the
  wake of a circular disk. Phys Fluids 27(6):064101, \doi{10.1063/1.4922109}

\bibitem[{Yassen(2005)}]{Yassen2005}
Yassen M (2005) Controlling chaos and synchronization for new chaotic system
  using linear feedback control. Chaos Solitons Fract 26(3):913 -- 920,
  \doi{10.1016/j.chaos.2005.01.047}

\bibitem[{Yu et~al(2006)Yu, Leeser, Tadmor, and Siegel}]{Yu2006}
Yu H, Leeser M, Tadmor G, Siegel S (2006) Real-time particle image velocimetry
  for feedback loops using {FPGA} implementation. J Aeros Comp Inf Com
  3(2):55--62, \doi{10.2514/1.18062}

\bibitem[{Yun et~al(2006)Yun, Kim, and Choi}]{Yun2006}
Yun G, Kim D, Choi H (2006) Vortical structures behind a sphere at subcritical
  {R}eynolds numbers. Phys Fluids 18(1):015102, \doi{10.1063/1.2166454}

\end{thebibliography}

\end{document}